\documentclass[10pt]{IEEEtran}

\IEEEoverridecommandlockouts
\usepackage{cite}
\usepackage{amsmath,amssymb,amsfonts,bm}
\usepackage{algorithmic}
\usepackage{graphicx}
\usepackage{textcomp}
\usepackage{xcolor}
\usepackage{amssymb}
\usepackage{subfigure}
\usepackage{graphicx,booktabs,multirow}
\usepackage[lined,boxed,commentsnumbered, ruled]{algorithm2e}
\usepackage{mathrsfs}
\usepackage[hidelinks,bookmarks=false]{hyperref}
\usepackage{pgfplots}
\usepackage{mathtools}
\usepackage{amsthm}
\usepackage{float}

\newcommand{\mini}{\operatorname{minimize}}
\newcommand{\maxi}{\operatorname{maximize}}
\newcommand{\subj}{\operatorname{subject~to}}

\newcommand{\diagg}{\operatorname{diag}}
\newcommand{\re}[1]{{\color{black}}{#1}}

\def\BibTeX{{\rm B\kern-.05em{\sc i\kern-.025em b}\kern-.08em
    T\kern-.1667em\lower.7ex\hbox{E}\kern-.125emX}}

\begin{document}

\title{Learning to Reflect and to Beamform for Intelligent Reflecting Surface with Implicit Channel Estimation
}

\author{
\IEEEauthorblockN{Tao~Jiang,  \IEEEmembership{Graduate Student Member,~IEEE}, Hei~Victor~Cheng, \IEEEmembership{Member,~IEEE}, and Wei~Yu, \IEEEmembership{Fellow,~IEEE}}
\thanks{Manuscript submitted on December 25, 2020, revised on March 15, 2021, and accepted on April 19, 2021 in IEEE Journal on Selected Areas in Communications.
The materials in this paper have been presented in part at the IEEE Global Communications
Conference (Globecom), December 2020 \cite{jiang2020learning}. This work was supported
by Huawei Technologies Canada and by Natural Sciences and Engineering Research Council
(NSERC) via the Canada Research Chairs program.  The authors are with The Edward S.~Rogers Sr.~Department of
Electrical and Computer Engineering, University of Toronto, Toronto, ON M5S 3G4, Canada
(e-mails: \{taoca.jiang@mail.utoronto.ca, hei.cheng@utoronto.ca, weiyu@ece.utoronto.ca\}).
}
}

\maketitle
\IEEEpeerreviewmaketitle

\begin{abstract}
Intelligent reflecting surface (IRS), which consists of a large number of tunable
reflective elements, is capable of enhancing the wireless propagation environment
in a cellular network by intelligently reflecting the electromagnetic waves from the base-station (BS) toward the users. The optimal tuning of the phase shifters at the IRS is,
however, a challenging problem, because due to the passive nature of reflective
elements, it is difficult to directly measure the channels between the IRS, the
BS, and the users.  Instead of following the traditional paradigm of first
estimating the channels then optimizing the system parameters, this paper
advocates a machine learning approach capable of directly optimizing both the
beamformers at the BS and the reflective coefficients at the IRS based on a
system objective. This is achieved by using a deep neural network to parameterize the mapping
from the received pilots (plus any additional information, such as the user
locations) to an optimized system configuration, and by adopting a \re{permutation
invariant/equivariant} graph neural network (GNN) architecture to capture the
interactions among the different users in the cellular network. Simulation results show that the proposed implicit channel estimation based approach is generalizable, can be interpreted, and can efficiently learn to maximize a sum-rate or minimum-rate objective from a much fewer number of pilots than the traditional explicit channel estimation based approaches.
\end{abstract}


\section{Introduction}
Conventional physical-layer communication system design has always been based
on the paradigm of first modeling the channel, then optimizing the transmitter
and the receiver design parameters according to the channel model. While this
conventional approach has served the communication engineers well for many
practical communication scenarios, the recent emergence of new communication
modalities involving passive reflectors for which channel estimation may not be
straightforward to perform has motivated the need for new approaches. This
paper investigates the design of reflective patterns for intelligent
reflecting surfaces (IRS), \re{also known as reconfigurable intelligent surface, } composed of a large number of tunable reflective
elements.  We advocate the use of machine learning techniques to bypass
explicit channel estimation and to directly design the beamforming and
reflective patterns to optimize a system-wide objective.

The promise of IRS stems from its ability to manipulate incident
electromagnetic waves toward the intended directions by controlling the phase
responses of the passive elements \cite{di2020smart,huang2020holographic}.
By intelligently adjusting these phases, in effect modifying the signal
propagation environment, the IRS can enhance the communication channel between
the transmitter and the receiver. In addition, due to its passive structure, the IRS requires very
little energy to produce the desired phase shifts for signals reflection.
Further, it can be flexibly integrated into various objects (e.g., walls,
ceilings), thus enabling a smooth deployment of IRS in existing wireless
communication networks \cite{di2020smart}. As a result, a wide range of
applications for IRS have been explored in the literature, e.g., for improving
network coverage \cite{subrt2012intelligent}, \re{boosting wireless spectral
efficiency \cite{guo2020weighted,yang2020mimo}, reducing power consumption of
data transmission \cite{wu2019intelligent,huang2019reconfigurable,zhu2020power}}, enhancing over-the-air computation performance \cite{jiang2019over} and enabling secure wireless communications
\cite{yu2020robust}.

This paper tackles the problem of how to optimally tune the IRS elements for
capacity enhancement in a multiuser cellular network. A conventional design
would have followed the approach of first estimating the channel, then
optimizing the design parameters. This is, however, not necessarily the best
approach for designing an IRS system, due to the following reasons. First,
the passive elements in the IRS have no ability to perform active signal
transmission and reception, so the channels to and from an IRS can only be
estimated indirectly.  Second, an IRS typically has a large number of
reflective elements, so the number of channel parameters to estimate can be
very large.  Third, conventional channel estimation is always with respect to
some artificial criterion (such as the mean squared error), which does not necessarily
correspond to the ultimate system objective. Finally, even if the channel is
perfectly known, the optimization of the beamformers and phase shifts is a
high-dimensional nonconvex problem, so finding an optimal solution
remains a difficult numerical problem.

The main idea of this paper is that a data-driven approach can be used to
overcome some of these difficulties. We show that by adopting a \emph{graph
neural network} (GNN) architecture that models the interaction between the IRS
and the multiple users in the system, it is possible to directly learn the
mapping from the received pilots to a desired set of beamformers at the
base-station (BS) and a desired reflective pattern at the IRS for maximizing a
system-wide objective, such as the sum rate or the minimum rate across the multiple
users. The GNN structure \re{has the properties that a permutation of the ordering of the users induces the same permutation on the BS beamformers but keeps the IRS reflective pattern fixed 
(known as \emph{permutation equivariant} and \emph{permutation invariant} properties, 
respectively)}. Such an architecture allows the possibility of generalizability  to networks with an arbitrary number of users. Numerical results show
that the proposed approach produces solutions that can be easily interpreted.
Overall, this paper shows that by bypassing the explicit channel estimation phase altogether, a
machine learning approach can achieve a significantly higher transmission rate
than the conventional channel estimation based approach, especially
in the limited pilot length regime.

\subsection{Related Works}

Many of the existing works on the optimization of IRS assume that perfect
channel state information (CSI) is available at the BS. Based on
the perfect CSI assumption, joint optimization of the IRS reflection and BS
beamforming can be carried out for different network objectives, e.g.,
minimizing the energy consumption \cite{wu2019intelligent}, maximizing the
spectral efficiency \cite{guo2020weighted}, or maximizing the minimum rate
\cite{alwazani2020intelligent}. In practice, CSI needs to be
estimated.  Due to the passive nature of the IRS elements, directly estimating
the CSI for IRS is not feasible. Instead, CSI estimation needs to be carried
out either at the BS or at the users based on the end-to-end reflected signals.
In this direction, \cite{mishra2019channel} proposes to solve the channel
estimation problem based on a binary reflection method by turning on the IRS
elements one at a time, \re{while \cite{wei2020channel} proposes a channel estimation method based on parallel factor decomposition for estimating the BS-IRS channel and the individual IRS-user channels.}  However, as the number of elements in an IRS is
typically quite large (in order to achieve higher beamforming gain
\cite{bjornson2019intelligent}), these channel estimation methods typically 
require a large pilot
overhead. To address this issue, \cite{zheng2019intelligent} proposes to group the
IRS elements into sub-surfaces, but at a cost of reduced beamforming
capability. More recently, \cite{chen2019channel} proposes a compressed sensing
based channel estimation method for the multiuser IRS-aided system, which reduces
the training overhead significantly but requires the assumption of channel
sparsity. Further, \cite{wang2019channel} proposes to reduce the training
overhead by exploiting the common reflective channels among all the users. All
these works fall into the paradigm of first estimating the channels from the
received pilot signals, then solving the reflection optimization problems based
on the estimated channels.

Recently, data-driven approaches have been introduced to address the challenges either in channel estimation or beamforming \cite{elbir2020survey}. For the channel estimation problem, \cite{liu2020deep} proposes a deep denoising neural network to enhance the performance of the model-based compressive channel estimation for mmWave IRS systems. The authors of \cite{elbir2020deep} propose a convolutional neural network to estimate both direct and cascaded channels from the received pilot signals through end-to-end training. Given perfect channels, the beamforming problem has been investigated from the perspective of the data-driven approach in \cite{gao2020unsupervised,feng2020deep,huang2020reconfigurable}.  In particular, \cite{gao2020unsupervised} proposes a multi-layer fully connected neural network to learn the phase shifts for a single-user system to reduce the algorithm run-time complexity of optimizing the IRS. Furthermore,  deep reinforcement learning is leveraged to optimize the phase shifts for the single-user system in \cite{feng2020deep}, and for the multiuser case in \cite{huang2020reconfigurable}.

This paper proposes to use a data-driven approach for optimizing the IRS.
The proposed approach is motivated by the success of using deep learning to
optimize wireless communication systems without explicitly estimating the channel
\cite{alkhateeb2018deep,cui2019spatial,huang2019indoor}. In particular,
\re{\cite{alkhateeb2018deep} shows that the beamforming vectors learned from
the received pilot signals can approach the achievable rate of the optimal
beamforming vectors for highly-mobile mmWave systems.} Further,
\cite{cui2019spatial} shows that based on the geographical locations of the
users, the deep learning approach is able to learn the optimal scheduling
without channel estimation.  Location information is also utilized in
\cite{huang2019indoor} to configure the IRS for indoor signal focusing using a
deep learning approach. In this paper, we primarily use the received pilots as
the input to the neural network, because the received pilot signal contains rich information
about both the large-scale and small-scale fading, but we also show that incorporating
the location information can further reduce the required pilot length.

\subsection{Main Contributions}

This paper casts the problem of designing the beamforming and reflective
patterns in an IRS system as a variational optimization problem whose
optimization variables are functionals, i.e., mappings from the received pilots to
the phase shifts at IRS and beamforming matrix at the BS.
We propose to parameterize this mapping using a neural network and to train the
neural network based on the training data to directly maximize a network utility
function.

While a fully connected neural network has already been shown to be able to
significantly reduce pilot length for sum-rate maximization in the conference
version of this paper \cite{jiang2020learning}, a further contribution of this
journal paper is that we propose a GNN architecture to better model the
interference among the different users in the network.
The proposed GNN is \re{permutation invariant/equivariant} across the users and therefore
provides better scalability and generalization ability.
For example, while a fully connected neural network would not have been able to
generalize when the number of users in the network changes (except by re-training a new neural network),
a GNN structure can easily have shared parameters across the components for different users, thereby
achieving generalizability.
It is worth noting that GNNs have been proposed to solve radio resource
allocation problems in \cite{eisen2020optimal,lee2019graph,shen2020graph}, but
these prior works all require perfect CSI and are not designed for IRS systems.

A key benefit of the data-driven approach for communication system design is that
it can easily incorporate different types of data as inputs to the neural network.
In this paper, we propose to incorporate the locations of the
users, so that the neural network can focus on learning
the small-scale fading component of the wireless channels. The numerical results
show that this significantly improves the utility maximization performance. We
remark that incorporating such heterogeneous information is not
easy to do in the conventional model-based approach.

This paper also shows that the same machine learning framework can be used to design systems under different overall network objectives. In particular, this paper shows the superior performance of the 
GNN for both the sum-rate maximization and the minimum-rate maximization problems in term of reducing the length of pilots required to achieve a target performance level.

A crucial requirement for the eventual adoption of the machine learning
approach to system-level optimization is the interpretability of its solutions.
Toward this end, this paper analyzes the beamforming patterns and the reflective
patterns that the GNN learns from training data. We show visually that the
deep learning approach is indeed performing the task of focusing the
electromagnetic waves toward the directions of the target users, hence
providing a level of confidence that the proposed approach is capable of finding an
optimized solution in the overall highly nonconvex optimization landscape.

To summarize, the main novelties and the key findings of this paper are as follows:
\begin{enumerate}
\item A deep neural network is used to directly map the received pilots to the optimized phase-shifts at the IRS and optimized beamforming matrix at the BS. By bypassing the explicit channel estimation stage, the proposed machine learning framework is able to utilize pilots more efficiently than the conventional channel estimation based approaches.
\item A GNN architecture is used to map the received pilots to the beamformers at \re{the} BS and the reflective pattern at \re{the} IRS. The GNN is \re{permutation invariant for the reflective pattern and permutation equivariant for the user beamformers}, thereby allowing generalizability to systems with different number of users.
\item The proposed neural network is able to incorporate heterogeneous inputs such as the location information of the users to further improve the system performance. 
\item The beamforming patterns obtained from the machine learning approach are visualized, showing that the proposed GNN is able to learn the correct beam focusing patterns for both the IRS and the BS.
\item Numerical simulations show that the proposed deep learning framework outperforms conventional model-based approaches for both the sum-rate maximization and the minimum-rate maximization problems. Further, the proposed GNN generalizes well to different signal-to-noise ratios (SNRs).
\end{enumerate}

We remark that a parallel and independent work \cite{ozdogan2020deep} has also
proposed the idea of using the received pilot signals to learn the
configuration of the phase shifts. However,  \cite{ozdogan2020deep} only
considers a single-user IRS system. Moreover, their method first uses an
alternating optimization approach to find an optimized set of beamforming
matrix and phase shifts, then uses a fully connected neural network to learn
the optimized solution in a supervised fashion.  In comparison, the method
proposed in this paper is more direct. Further, our method is designed for the
multiuser setting and can be more easily generalized to systems with different
number of users.

\subsection{Organization of the Paper and Notations}

The rest of the paper is organized as follows.
Section \ref{sec:system_model} describes the system model and problem formulation.
Section \ref{sec:baseline} describes the uplink pilot transmission and the conventional channel estimation strategy.
Section \ref{sec:deep_learning_framework} describes the proposed deep learning framework and the GNN architecture for the IRS system.
Sections \ref{sec:sum_rate} and \ref{sec:maxmin} provide simulation results for
the proposed scheme for both sum-rate and min-rate objectives.
Section \ref{sec:interpretation} interprets the results by visualizing the beamforming and reflective patterns. Finally, conclusions are drawn in Section \ref{sec:conclusion}.

The notations used in this paper are as follows. Lower case letters are used to denote scalars.
Lower case bold-faced letters are used to denote column vectors. Upper case bold-faced letters
are used to denote matrices. We use $[\cdot]_i$ to denote an element of a vector and $(\cdot)^\top$ and $(\cdot)^{\sf H}$ to denote transpose
and Hermitian transpose of matrices.  We use $\mathcal{CN}(\cdot,\cdot)$ to denote a complex Gaussian distribution,
and $\mathbb{E}[\cdot]$ to denote expectation.

\section{Problem Formulation}
\label{sec:system_model}

\subsection{System Model}
\begin{figure}[!t]
    \centering
    \includegraphics[width=3.5in]{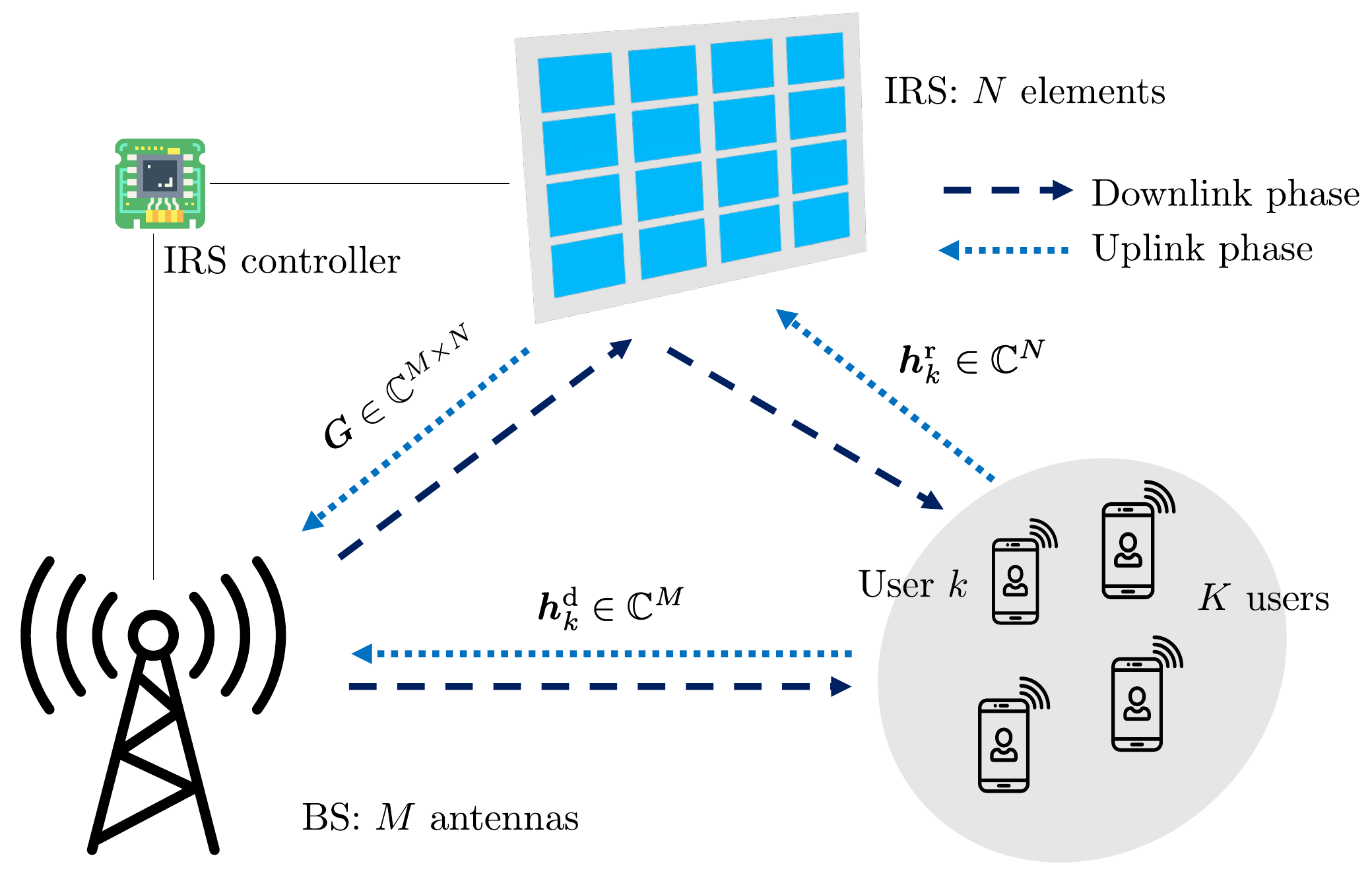}
    \caption{IRS assisted  \re{multiuser MISO} system}
    \label{fig:system_model}
\end{figure}

Consider an IRS assisted downlink  \re{multiuser MISO} system where $K$ single-antenna
users are served by a BS with $M$ antennas. An IRS equipped with an array of $N$
passive reflective elements is deployed to aid the transmission between the BS
and users. The BS controls the IRS through an IRS controller, which is capable
of adjusting the phases of the array elements in order to reflect the incident
signals to desired directions, as shown in Fig.~\ref{fig:system_model}.

We use $\bm G\in\mathbb{C}^{M\times N}$ to denote
the uplink channel matrix from the IRS to the BS, $\bm h_k^{\rm r}\in\mathbb{C}^{N}$  to denote
the uplink channel vectors from user $k$ to the IRS, and $\bm h_k^{\rm d}\in\mathbb{C}^{M}$ to denote
the uplink channel vectors from user $k$ to the BS.  We assume channel
reciprocity, so that the channel matrices in the downlink direction are the transpose of
the uplink channels.

Let $s_k\in\mathbb{C}$ be the symbol to be transmitted from the BS to user $k$. The BS uses a beamforming
strategy to transmit $s_k$ through a beamforming vector $\bm w_k\in\mathbb{C}^{M}$,
where the beamformers must satisfy a power constraint
$\sum_{k=1}^K\mathbb\|\bm w_k\|_2^2\le P_d$.
Let $\bm v=[e^{j\omega_1},e^{j\omega_2},\cdots,e^{j\omega_N}]^\top$ be the reflection coefficients at the IRS, where  $\omega_i\in [-\pi ,\pi)$ is the phase shift of the $i$-th element.
Then, the received signal $ r_k $ at the user $k$ is given by
\begin{align}
    r_k &= \sum_{j=1}^K(\bm h_k^{\rm d}+ \bm G\diagg(\bm v)\bm h_k^{\rm r})^\top\bm w_j s_j + n_k\\
        &= \sum_{j=1}^K(\bm h_k^{\rm d}+ \bm A_k\bm v )^\top\bm w_j s_j + n_k,
\end{align}
where $\bm A_k = \bm G\diagg(\bm h_k^{\rm r}) \in\mathbb{C}^{M\times N}$ is the
cascaded channel between the user $k$ and the BS through reflection at the IRS,
and $ n_k\sim\mathcal{CN}( 0,\sigma_0^2)$ is the additive white Gaussian noise.

A block-fading model is assumed in which the channel coefficients remain constant
during a coherence block, but change independently from block to block.
An achievable rate $R_k$ for user $k$ can be computed as
\begin{equation}
    R_k = \log\left(1+\frac{|(\bm h_k^{\rm d}+ \bm A_k\bm v )^\top\bm w_k|^2}{\sum_{i=1,i\neq k}^K |(\bm h_k^{\rm d}+ \bm A_k\bm v )^\top\bm w_i|^2+\sigma_0^2}\right),
\end{equation}
where multiuser interference is treated as noise.
The beamforming vectors $\bm w_k$'s at the BS and phase shifts $\bm v$ at the
IRS can be jointly optimized to maximize a network utility $U(R_1 \ldots R_K)$,
which is a function of the achievable rates of all the users. Common utility functions
include the sum rate $\sum_{k=1}^{K} R_k$, and the minimum rate $\min_k R_k$.

To optimize the transmit beamformers at the BS and the reflection coefficients
at the IRS, the knowledge of the cascaded channel matrix $\bm A_k$ and the
channel vector $\bm h_k^{\rm d}$ for $k=1,\cdots,K$ is needed either explicitly or
implicitly. Toward this end, a pilot transmission phase is used before the data
transmission phase to gain knowledge of $\bm A_k$'s and $\bm h_k^{\rm d}$'s.
By uplink-downlink channel reciprocity, channel estimation can take place in
the uplink \cite{mishra2019channel}. Specifically, we use an uplink
pilot transmission phase, in which the user $k$ sends a pilot sequence, $x_k(\ell)$,
$\ell = 1,\cdots,L$, which is reflected through the IRS and received at the BS.
The received signal $\bm y(\ell)$ at the BS in the time slot $\ell$ can be expressed as
\begin{align}\label{eq:uplink}
    \bm y(\ell) &=\sum_{k=1}^{K}(\bm h_k^{\rm d}  +  \bm G \diagg(\bm v(\ell)) \bm h_k^{\rm r}) x_k(\ell) +\bm n(\ell) \\
    &=\sum_{k=1}^K(\bm h_k^{\rm d}+\bm A_k\bm v(\ell))x_k(\ell)+\bm n(\ell), 
\end{align}
where diag$(\bm v(\ell))$ is a diagonal matrix with $\bm v(\ell)$, the phase shifts of
IRS in time slot $\ell$ of the pilot phase, on its diagonal, and
$\bm n(\ell)\sim\mathcal{CN}(\bm 0,\sigma_1^2\bm I)$ is the additive Gaussian noise. 
Note that there are $(M+N)K+MN$ unknown channel coefficients in $\bm G$, $\bm h_k^{\rm d}$'s and $\bm h_k^{\rm r}$'s, for $k=1,\cdots,K$. Since the number of elements $N$ in a typical IRS is generally large (possibly in the hundreds), it is challenging to estimate the channels when the pilot length is short. 

\subsection{Problem Formulation}
The main idea of this paper is that since the final goal is to optimize the rates $R_k$'s, instead of explicitly estimating all the channel coefficients as an intermediary step, we can exploit the pilot phase more efficiently by mapping the received pilots directly to the optimized transmission strategy for utility maximization, in effect, bypassing channel estimation.

To this end, we propose to design the optimal beamforming vector $\bm w_k$'s and the reflection phase shifts $\bm v$ based directly on the received pilots $\bm{Y} = [\bm y(1),\bm y(2),\cdots,\bm y(L)]\in\mathbb{C}^{M\times L}$. Specifically, given the matrix  $\bm{Y}$, our goal is to solve the following optimization problem
\begin{equation}\label{prob:formulation}
    \begin{aligned}
        & \underset{\begin{subarray}{c}
		(\bm W, \bm v)= g(\bm{Y})
        \end{subarray}}{\maxi}&\quad &\mathbb{E}\left[U( R_1(\bm v,\bm W),\ldots, R_K(\bm v,\bm W) )\right] \\
        & \subj&\quad& \sum_k\mathbb\|\bm w_k\|^2\le P_d,\\
        &&&|v_i| = 1,i=1,2,\cdots,N,
    \end{aligned}
\end{equation}
where $\bm W=[\bm w_1,\cdots,\bm w_k]$ is the beamforming matrix at BS, $g(\cdot)$ is a function that maps the received pilots to the beamforming matrix $\bm W$ and the phase shift vector $\bm v$, and $U(\cdot)$ is the network utility function. The expectation here is over the random channel realizations and the noise in the uplink pilot transmission phase. In the rest of the paper, we use the sum-rate maximization and the max-min fairness problem as examples to illustrate our idea, but the proposed method can be extended to other utility functions as well.

Solving problem \eqref{prob:formulation} is computationally challenging, because it is a variational optimization problem with a nonconvex objective
function. To tackle this problem, we propose to parameterize the mapping
function $g(\cdot)$ by a deep neural network, and to learn the parameters of the neural network from data. This is motivated by the universal approximation property of the neural networks \cite{hornik1989multilayer}.
Before presenting the proposed approach, we first discuss the uplink pilot transmission stage and the conventional approach of channel estimation followed by network utility maximization for solving the
problem \eqref{prob:formulation}. 

\section{Uplink Pilot Transmission and Conventional Channel Estimation}
\label{sec:baseline}
In this section, we describe the conventional approach to solving the problem \eqref{prob:formulation}, which consists of an uplink channel estimation phase and a downlink utility maximization phase. Given the estimated channels, the downlink utility maximization problem is well investigated. For instance, the downlink sum-rate maximization problem can be solved using the algorithm proposed in \cite{guo2020weighted}, and the minimum rate maximization problem is studied in \cite{alwazani2020intelligent}. Below we focus on the uplink pilot transmission and the channel estimation step.

\subsection{Uplink Pilot Transmission}
We adopt the pilot transmission strategy proposed in \cite{chen2019channel} to design the uplink pilots and the phase shifts at the IRS in the pilot phase. 
In particular, the total training slots $L$ is divided into $\tau$ sub-frames, each of which consists of $L_0 = K$ symbols (i.e., $L=\tau L_0$).
The users simultaneously send their pilot sequences
$\bm x_k^{\sf H} = [x_k(1),x_k(2),\cdots,x_k(L_0)]$ of length $L_0$ to the BS, repeated over
$\tau$ sub-frames.
The pilot sequences of all users are designed to be orthogonal to each other so that they can be
decorrelated at the BS, i.e.,  $\bm x_{k_1}^{\sf H}\bm x_{k_2}=0$ if $k_1 \neq
k_2$ and $\bm x_{k_i}^{\sf H}\bm x_{k_i}=L_0 P_u$ where $P_u$ is the uplink
pilot transmission power. In the meanwhile, the IRS keeps the phase shifts fixed
within each sub-frame, but uses different phase shifts in different sub-frames so that
both the users-to-IRS and the IRS-to-BS channels can be measured.

The BS decorrelates the received pilots in each sub-frame by matching the pilot
sequence for each user.  Let $\bm{\bar{Y}}(t) = [\bm y((t-1)L_0+1),\cdots,\bm y(tL_0)]$
denote the received pilots in sub-frame $t$.  Let $\bm{\bar v}(t)$ be the phase
shifts at the IRS in sub-frame $t$. Then, $\bm{\bar{Y}}(t)$ can be expressed as:
\begin{align}
	\bm{\bar{Y}}(t) &=\sum_{k=1}^K(\bm h_k^{\rm d}+\bm A_k\bm{\bar{v}}(t))\bm x_k^{\sf H}+\bm{\bar{N}}(t), \quad t=1,\cdots,\tau,
\end{align}
where $\bm{\bar{N}}(t)$ is a noise matrix with each column independently distributed as $\mathcal{CN}(\bm 0, \sigma_1^2\bm I)$. By the orthogonality of the pilots, we can form $\bm{\bar{y}}_{k}(t)\in\mathbb{C}^M$, i.e., the contribution from user $k$ at the $t$-th sub-frame, as given in \cite{chen2019channel}:
\begin{align}
	\bm{\bar y}_{k}(t) &= \frac{1}{L_0} \bm{\bar Y}(t)\bm x_k = \bm h_k^{\rm d}+\bm A_k\bm{\bar{v}}(t)+\bm{\bar{n}}(t)\\
    &\triangleq\bm F_k \bm q(t)+\bm{\bar{n}}(t),
\end{align}
where $\bm{\bar{n}}(t) = \frac{1}{L_0}\bm{\bar{N}}(t)\bm x_k$, and the combined channel matrix is defined as $\bm F_k \triangleq [\bm h_k^{\rm d},\bm A_k]$ and the combined phase shifts is defined as $\bm q(t)\triangleq[1,\bm{\bar{v}}(t)^\top]^\top$.
Recall that we have $\tau$ sub-frames in total. Then, by denoting $\bm {\tilde Y}_k = [\bm{\bar{y}}_{k}(1),\cdots,\bm{\bar{y}}_{k}(\tau)]$ as a matrix of received pilots across $\tau$ sub-frames, we have
\begin{align}\label{eq:linear_sys}
    \bm{\tilde Y}_{k} =\bm F_k\bm Q + \bm{\tilde{N}},
\end{align}
where $\bm Q = [\bm q(1),\cdots,\bm q(\tau)]$  and $\bm{\tilde{N}}=[\bm{\bar{n}}(1),\cdots,\bm{\bar{n}}(\tau)]$. 

The channel estimation problem aims to estimate the combined matrix $\bm F_k$ for $k=1,\dots,K$.
Typically, to ensure that the matrix $\bm Q$ is full rank so that $\bm F_k$ can be recovered successfully,  we need at least $\tau=N+1$, i.e., a total $(N+1)K$ pilot symbols are needed.
When $\tau=N+1$, one choice of $\bm Q$ is a DFT matrix as suggested in \cite{zheng2019intelligent}. For comparison purposes, we also consider the more general case where $\tau \neq N+1$. In this case, $\bm Q$ is not a square matrix. One way to construct $\bm Q $ is to first form a $d\times d$ DFT matrix $\bm Q^\prime$ with $d=\max(\tau,N+1)$, then truncate $\bm Q^\prime$ to the first $\tau$ columns or the first $N+1$ rows. Another possibility is to independently construct vectors $\bm{\bar{v}}(t), ~t=1,\cdots,\tau $, randomly. Specifically, the phase of $[\bm{\bar{v}}(t)]_i $ can be constructed by drawing from a uniform random variable in $[-\pi, \pi)$. In this paper, we use the second approach to construct $\bm Q $ if $\tau<N+1 $, and use the first approach in other cases.
These are empirically good choices.

\subsection{Conventional Channel Estimation}
To estimate the channel matrix $\bm F_k$ from \eqref{eq:linear_sys}, we can use the minimum mean-squared error (MMSE) estimator, obtained by solving the following problem
\begin{equation}\label{prob:lmmse}
    \begin{split}
        \underset{f(\cdot)}{\mini}\quad &\mathbb{E}\left[\| f(\bm{\tilde Y}_k) -\bm F_k\|_F^2\right].
    \end{split}
\end{equation}
The optimal solution to problem \eqref{prob:lmmse} is given by 
\begin{align}
    f(\bm {\tilde Y}_k) = \mathbb{E}[\bm F_k|\bm {\tilde Y}_k].
\end{align}
However, for general channel fading distributions, the optimal solution is computationally intensive to implement. A low-complexity approach is to constrain the estimator  $f(\cdot)$ to be linear, which results in the linear MMSE (LMMSE) method. When the rows of $\bm F_k$ and the rows of ${\bm{\tilde{N}}}$ are i.i.d., the LMMSE estimator is as follows \cite{bjornson2009framework}
\begin{IEEEeqnarray}{rCl}
    \hat{\bm F_k} & = &(\bm {\tilde Y}_k- \mathbb{E}[\bm {\tilde Y}_k])\left(\mathbb{E}[(\bm {\tilde Y}_k-\mathbb{E}[\bm {\tilde Y}_k])^{\sf H}(\bm {\tilde Y}_k-\mathbb{E}[\bm {\tilde Y}_k])] \right) ^{-1}\notag\\
    & & \quad \mathbb{E}[(\bm {\tilde Y}_k-\mathbb{E}[\bm {\tilde Y}_k])^{\sf H} (\bm F_k-\mathbb{E}[\bm F_k])]+\mathbb{E}[\bm F_k].
\end{IEEEeqnarray}
The estimates of $\bm h^{\rm d}_k, \bm A_k$ can then be obtained from $\hat{\bm F_k}$. Note that the LMMSE estimator is an optimal solution to \eqref{prob:lmmse} only if the unknown $\bm F_k$ is Gaussian distributed.  We note that this LMMSE channel estimation method for  \re{multiuser MISO} system is also proposed in  \cite{alwazani2020intelligent}.

\begin{figure*}[!t]
    \centering
    \includegraphics[width=0.9\linewidth]{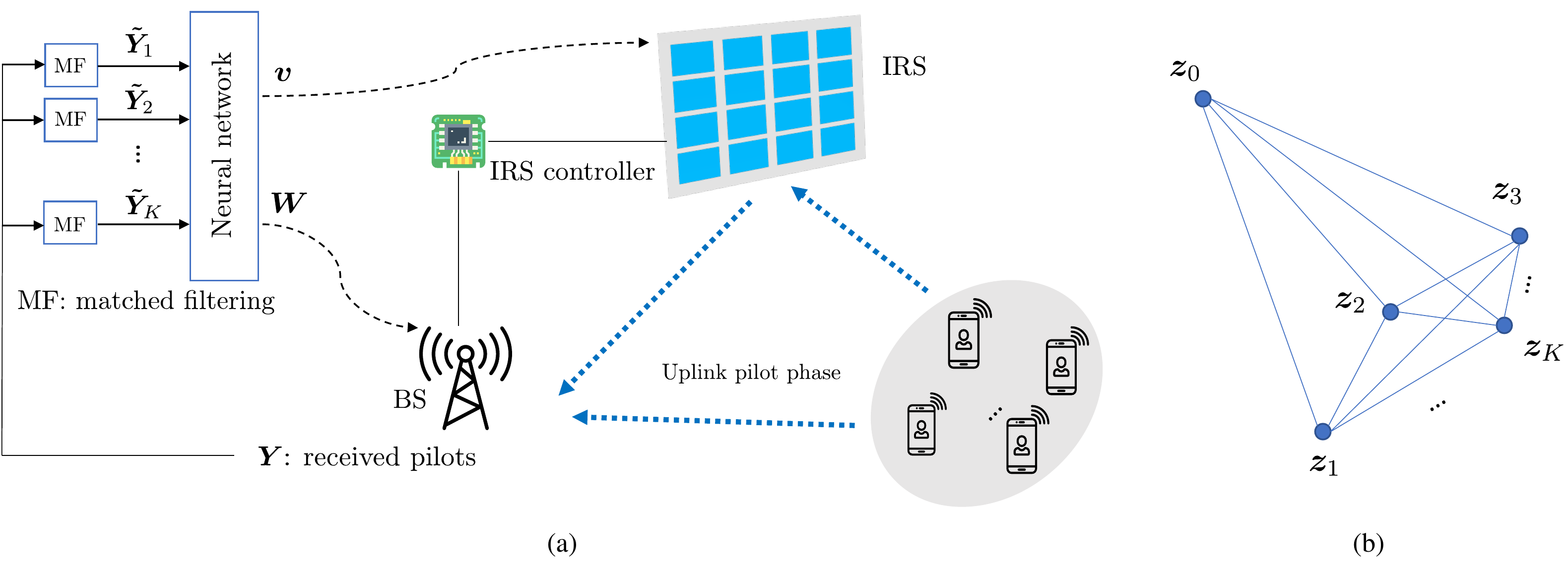}
	\caption{(a) Proposed deep learning framework for directly designing the beamformers and phase shifts based on the received pilots; (b) Graph representation of the network. The node representation vector $\bm z_0$ corresponds to the IRS, and $\bm z_1,\cdots,\bm z_K$ correspond to the users.}
    \label{fig:dnn_overall_arch}
\end{figure*}

\section{Proposed Deep Learning Framework}
\label{sec:deep_learning_framework}

The conventional channel estimation approach aims to solve \eqref{prob:lmmse}, i.e.,
to recover entries of $\bm F_k$ given the received pilots $\bm {\tilde Y}_k$ using a
mean squared error metric. However, recovering the channels is not the
goal. Our ultimate objective is to maximize the network utility as in
\eqref{prob:formulation}.  The main idea of this paper is to bypass explicit channel estimation and to solve
problem \eqref{prob:formulation} directly.

Specifically, we aim to use a neural network to represent the mapping
function $g(\cdot)$ in problem \eqref{prob:formulation}, and
to pursue a data-driven approach to train the neural network so that it mimics
the optimal mapping from the received pilot signals to the beamformers and
the phase shifts for network utility maximization. This overall framework is
depicted in Fig.~\ref{fig:dnn_overall_arch}(a). In this section,
we describe the neural network architecture suited for this task.

\subsection{Graphical Representation of Users and IRS}
A central task in a multiuser cellular network is the management of the
interference between the users. Toward this end, the beamformers at the BS and
the phase shifts at the IRS must be coordinated so that the mutual interference is
minimized. This paper proposes to use a neural network architecture, called
GNN, which is based on a graph representation of the beamformers for the users
and the phase shifts at IRS, to capture the multiuser interference.
The graph consists of $K+1$ nodes as shown in Fig.~\ref{fig:dnn_overall_arch}(b). The IRS is represented by node 0 and the $K$ users are represented by nodes $1$ to $K$.
A representation vector, denoted as $\bm z_k$, $k=0,1,\cdots,K$, is associated
with each node. \re{The goal is to encode all the useful information about each
corresponding node in these representation vectors. The representation vectors 
are updated layer by layer in a GNN, taking into account all the representation 
vectors in the previous layer as input. After multiple
layers, the representation vector of each node would contain sufficient information
for designing the beamforming vectors and the reflection coefficients.
Specifically, the} GNN is trained in such a way so that the phase shifts of the
IRS can be subsequently obtained from $\bm z_0$, and the beamforming matrix at
the BS can be obtained from ${\bm z_1},\cdots,{\bm z_K}$.

As compared to a fully connected neural network, the GNN more naturally captures
the interactions between the users and the IRS.  By explicitly embedding these
interactions into the neural network architecture, the GNN is better
able to learn a mechanism for reducing the interference between users. In particular, the update of each user node is a function of all its neighboring user nodes and
the IRS node, which enables the GNN to learn to avoid interference. The update of the
IRS node is a function of all the user nodes, which enables the GNN to learn to
configure the phase shifts to spatially separate the channels for all the users.

A useful feature of the GNN is that it is able to capture the \re{permutation
invariant and permutation equivariant properties}  \cite{eisen2020optimal,shen2020graph}
of the network utility maximization problem \eqref{prob:formulation}.
That is, if we permute the index labels
of the users in the problem, the neural network should
output the same set of beamforming vectors $\bm w_k$ with permuted indices
and the same phase shifts $\bm v$.  \re{Here, \emph{permutation invariance} means that the phase shifts $\bm v$ is independent of the ordering of the user channels, and \emph{permutation equivariance} means that if the user channels are permuted, the beamforming vectors $\bm w_k $ would be permuted in the same way.} 
These properties are not easy to learn in a fully connected neural network, but
are naturally embedded in the GNN.

By tailoring to the problem structure, the GNN also reduces the model complexity as
compared to the fully connected neural network. More importantly, the
parameters of the GNN can be tied across the users, so that it can be easily
generalized to scenarios with different number of users. This is in contrast to
fully connected neural networks, whose parameters need to scale with the number
of users, which makes it difficult to generalize.

\subsection{GNN Architecture}
\begin{figure*}[t]
    \centering
    \subfigure{\includegraphics[width=0.95\linewidth]{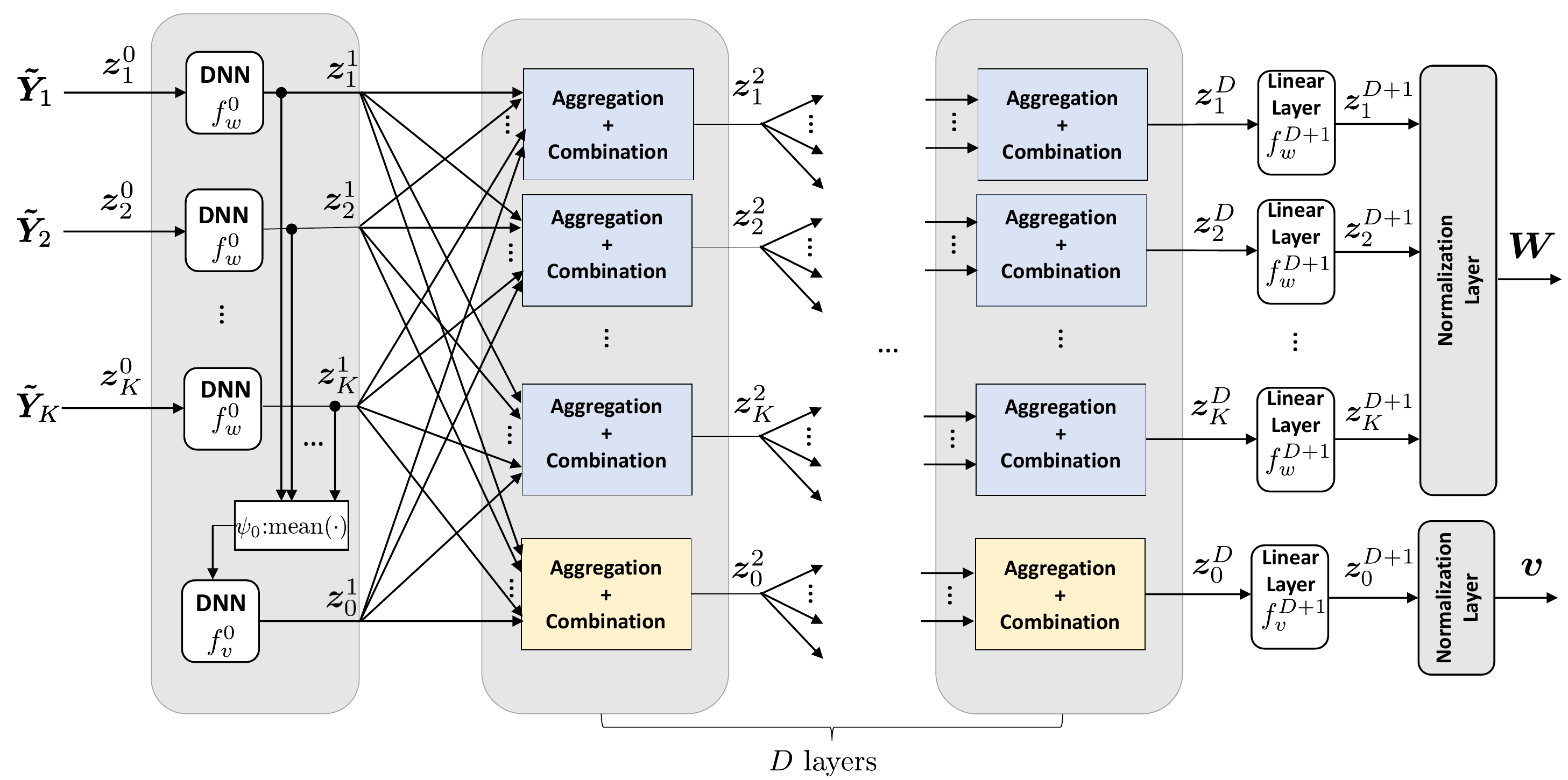}}
    \caption{Overall graph neural network architecture with an initialization layer, $D$ updating layers, and a final normalization layer.}\label{fig:dnn}
    \vspace{1em}
    \subfigure[The IRS node.]{\includegraphics[width=2.6in]{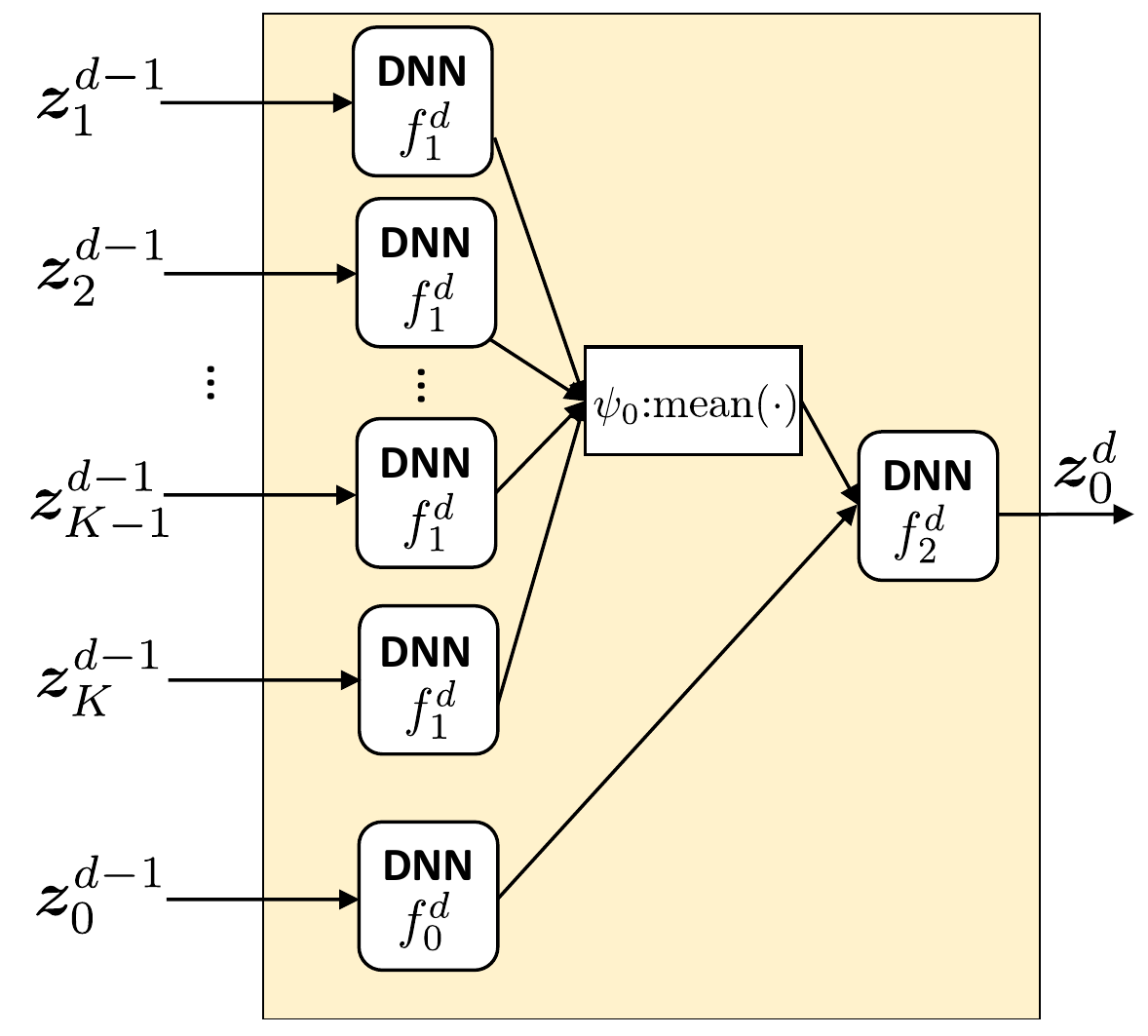}\label{fig:gnn_nodes2}}\quad
    \subfigure[The user node $k$.]{\includegraphics[width=2.6in]{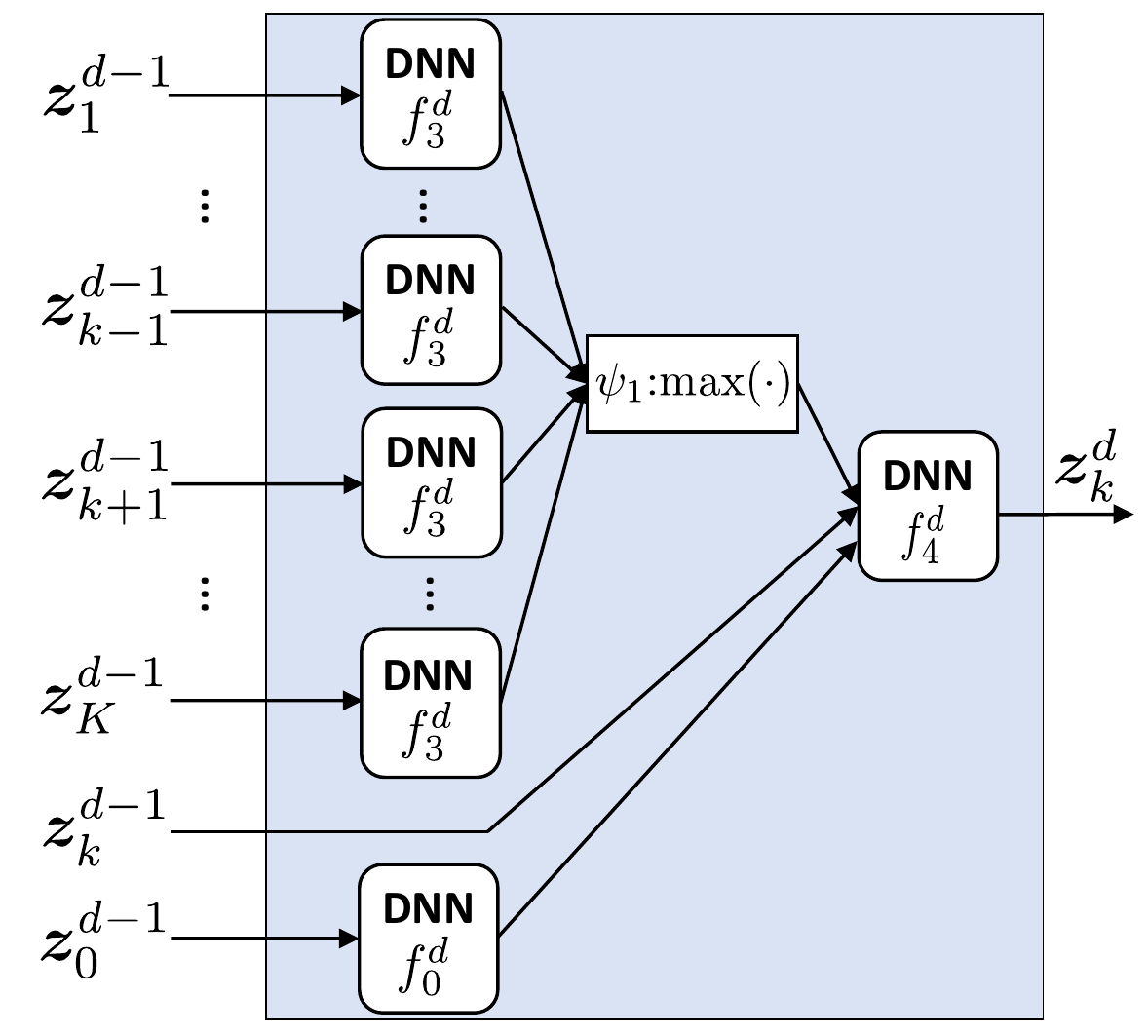}\label{fig:gnn_nodes1}}
    \caption{Aggregation and combination operations of the $d$-th layer for (a) the IRS node, and (b) the user nodes.}
\end{figure*}

We now describe the proposed GNN architecture and the training process.
The overall GNN aims to learn the graph representation vector $\bm z_k$ through an
initialization layer, $D$ aggregation and combination layers, and a final layer
that includes normalization.  It takes the input feature from the user node of the graph
as the initial value of the representation vector, denoted as $\bm z_k^0$, then
updating them through the $D$ layers to produce $\bm z_k^d$, $d=1,\cdots,D$, and
finally a linear layer to produce $\bm z_k^{D+1}$, which is mapped to the beamformer
matrix $\bm W$ and the phase shifts $\bm v$ via normalization.  The overall
architecture is shown in Fig.~\ref{fig:dnn}.

\subsubsection{Initialization Layer}

The initialization layer takes the input features from the user nodes as the
initial representation vector $\bm z_k^0$, $k=1,\cdots,K$, then trains one layer
of the neural network to produce $\bm z_k^1$, $k=0,\cdots,K$ for the subsequent
layers. Note that the IRS node does not have input features, because only the users transmit pilots.
 
The input features from the user nodes are simply the received pilots, i.e.,
vectorized form of the matrix $\bm {\tilde Y}_k$ with real and imaginary components separated:
\begin{align}
	\bm z_k^0 = [\operatorname{vec}(\Re\{\bm{\tilde{Y}}_k\})^\top, \operatorname{vec}(\Im\{\bm{\tilde{Y}}_k\})^\top]\re{^\top .}
\end{align}
As mentioned earlier, it is easy for the neural network to incorporate additional
useful information about each user into the input feature vector $ \bm z_k^0$.
For example, if the locations of the users are available, the input feature
vector $\bm z_k^0$ can be
\begin{align}
    \bm z_k^0 = [\operatorname{vec}(\Re\{\bm{\tilde{Y}}_k\})^\top, \operatorname{vec}(\Im\{\bm{\tilde{Y}}_k\})^\top, \bm{l}_k^\top ]^\top,
\end{align}
where $\bm{l}_k$ is the three-dimensional vector denoting the coordinates of the location of
the user $k$.  

Given the input feature vector $\bm z_k^0 $, we use a layer of fully connected neural networks, denoted as $f_w^0(\cdot)$,  to produce $\bm z_k^1$ for the user nodes, i.e.,
\begin{equation}
    \bm z_k^1 = f_w^0(\bm z_k^0), \qquad k=1,\cdots,K.
\end{equation}
For the IRS node, we take inputs from all the user nodes and process them using
a permutation invariant function $\psi_0(\cdot)$ first, then a fully connected neural network $f_v^0(\cdot)$ as
\begin{equation}
    \bm z_0^1 =f_v^0\left( {\psi_0}\left( \bm z_1^{0},\cdots, \bm z_K^{0}\right)\right). \label{eq:update_z0}
\end{equation}
In the implementation, we choose $\psi_0$ as the element-wise mean function,
i.e.,
\begin{equation}
[\psi_0(\bm z_1^0,\cdots,\bm z_K^0)]_i = \frac{1}{K}\sum_{k=1}^K [\bm z_k^0]_i.
\label{eq:element_mean}
\end{equation}
This is a reasonable choice, because all the pilots are reflected through the IRS,
so all the received pilots contain equal amount of information about IRS.

The representation vectors $(\bm z_0^1, \bm z_1^1, \cdots, \bm z_K^1)$ now contain
features about the IRS and the user channels, respectively, and are subsequently
passed to the $D$ updating layers of the GNN in order to eventually produce the IRS
reflective coefficients $\bm v$ from $\bm z_0^D$ and the user beamformers $\bm w_k$
from $\bm z_k^D$.

\subsubsection{Updating Layers}

The update of the  representation vector $ \bm z_k$ in the $d$-th layer is based on combining its previous representation and the aggregation of the representations from its neighboring nodes. In a general GNN, this is given by \cite{xu2019powerful}
\begin{align}\label{eq:gnn_update_0}
    \bm z_k^{d} =f_{\operatorname{combine}}^{d} \left(\bm z_k^{d-1}, f_{\operatorname{aggregate}}^{d} \left(\{\bm z_j^{ d-1}\}_{j\in\mathcal{N}(k)} \right)\right),
\end{align}
where $\mathcal{N}(k)$ denotes the set of neighboring nodes of the node $k $, $f_{\operatorname{aggregate}}^{d}(\cdot)  $ and $f_{\operatorname{combine}}^{d}(\cdot)$ are the aggregation function and combining function of the layer $d$, respectively.

A key in designing the GNN is to choose a suitable aggregation function $f_{\operatorname{aggregate}}^{d}(\cdot)$ and combining function $f_{\operatorname{combine}}^{d}(\cdot)$ in \eqref{eq:gnn_update_0} so that the GNN is scalable and generalizes well.  An efficient implementation of $f_{\operatorname{aggregate}}^{d}(\cdot) $ takes the following form \cite{fey2019fast,shen2020graph}:
\begin{align}
    f_{\operatorname{aggregate}}^{d}\left(\{\bm z_j^{ d-1}\}_{j\in\mathcal{N}(k)} \right) = 	\psi\left(\{f_{\operatorname{nn}}^d(\bm z_j^{ d-1})\}_{j\in\mathcal{N}(k)} \right),
\end{align}
where $\psi(\cdot)$ is a function that is invariant to the permutation of the inputs, e.g., element-wise max-pooling or element-wise mean pooling, and $f_{\operatorname{nn}}^d(\cdot)$ is a fully connected neural network. In addition, the combining function $f_{\operatorname{combine}}^{d}(\cdot)$ in  \eqref{eq:gnn_update_0} can also be implemented by a fully connected neural network for complicated optimization problems \cite{fey2019fast,shen2020graph}.

We adopt this framework to design the GNN for solving the problem \eqref{prob:formulation}.  We should note that our problem requires permutation \re{equivariance} with respect to the user nodes \re{and permutation invariance with respect to} the IRS node, so the aggregation and combination operations for the IRS and the \re{user} nodes are designed differently.

For the IRS node, we derive ${\bm z}_0^{d}$, the node representation vector in
the ${d}$-th layer from the previous layer as follows:
\begin{align}\label{eq:gnn_update_2}
    {\bm z}_0^{d} = {f_2^{d}}\left({f_0^{d}}(\bm z_0^{d-1}), {\psi_0}\left(f_1^{d}(\bm z_1^{d-1}),\cdots,f_1^{d}(\bm z_K^{d-1}) \right)\right),
\end{align}
where $f_0^{d}(\cdot)$, $f_1^{d}(\cdot)$ and $f_2^{d}(\cdot)$
are fully connected neural networks, and the aggregation function $\psi_0(\cdot) $ is chosen to be the element-wise mean function over the users as in \eqref{eq:element_mean}, which performs well empirically and corresponds to the fact that the IRS reflective pattern needs to serve all users.

For the user nodes, we recognize that aggregation should be with respect
to all the other user nodes excluding the IRS node and propose the update equation for
the node representation vectors $\bm z_k^d$ for $k=1,\cdots,K$ in the $d$-th layer
to take the form
\begin{align}\label{eq:gnn_update_1}
    &{\bm z}_k^{d} = f_4^{d}\left({f_0^{d}}(\bm z_0^{d-1}),{\bm z}_k^{d-1}, {\psi_1}\left(\left\{f_3^{d}(\bm z_j^{d-1})\right\}_{\forall {j\neq 0,j\neq k}}\right)\right),
\end{align}
where $f_3^{d}(\cdot)$, $f_4^{d}(\cdot)$ are fully connected neural networks and  ${\psi_1}(\cdot)$ is chosen to be the element-wise max-pooling function,
\begin{equation}
[\psi_1(\bm z_1,\cdots,\bm z_K)]_i = \max([\bm z_1]_i,\cdots,[\bm z_K]_i),
\end{equation}
which performs well empirically and corresponds to the fact that multiuser interference
is typically dominated by the strongest user.

The update of the node representation vectors in \eqref{eq:gnn_update_2} and \eqref{eq:gnn_update_1} is shown in Fig.~\ref{fig:gnn_nodes2} and Fig.~\ref{fig:gnn_nodes1}, respectively.
\re{We remark that there are many different design choices for the aggregation and combination operations. There is no general theory on how to choose these permutation invariant functions. In most works in the literature, the GNN architectures are designed based on empirical trials. In the simulation section of this paper, we shall see that the adopted GNN framework and the choice of the permulation invariant functions can achieve excellent performance.}

\subsubsection{Normalization layer}

After $D$ update layers, the representation vectors produced by the GNN, i.e.,
$\bm z_k^{D},~ k=0,\cdots,K $, are passed to a normalization layer to produce
the reflective coefficients $\bm v \in \mathbb{C}^{N}$ and
the beamforming matrix $\bm W \in \mathbb{C}^{M\times K}$,
while ensuring that the unit modulus constraints on $\bm v $ and the total power
constraint on $\bm W$ are satisfied.

To this end, we first take $\bm z_0^{D} $
as the input to a linear layer $f_v^{D+1}(\cdot)$ with $2N$ fully connected units
as follows:
\begin{align}
    &\bm z_0^{D+1} =f_v^{D+1} (\bm z_0^{D})\in \mathbb{R}^{2N}.\label{eq:normalize_v1}
\end{align}
Then a normalization layer outputs the real and imaginary components of
the reflection coefficients $\bm v$ as follows:
\begin{align}
    &\bm Z_v = [\bm z_0^{D+1}(1:N), ~ \bm z_0^{D+1}(N+1:2N)]\in \mathbb{R}^{N\times 2}, \label{eq:normalize_v2}\\
    &v_i =\frac{[\bm Z_v]_{i1}}{\sqrt{[\bm Z_v]_{i1}^2 + [\bm Z_v]_{i2}^2}}+j\frac{[\bm Z_v]_{i2}}{\sqrt{[\bm Z_v]_{i1}^2 + [\bm Z_v]_{i2}^2}} , ~\forall i,\label{eq:normalize_v3}
\end{align}
where $[\bm Z]_{ik}$ denotes the element in the $i$-th row and $k$-th column of the matrix $\bm Z$, and the notation $\bm z(i_1:i_2)$ denotes the subvector of $\bm z $ indexed from $i_1$ to $i_2$.

Similarly, to output the beamforming matrix, we first pass $\bm z_k^{D}$ through a linear layer $f_w^{D+1}(\cdot) $ with  $2M$ units
\begin{align}
    &\bm z_k^{D+1} =f_w^{D+1} (\bm z_k^{D})\in \mathbb{R}^{2M},~k=1,\cdots,K,
\end{align}
then use the following normalization steps to produce the beamforming matrix $\bm W$:
\begin{align}
    &\bm Z_w = [\bm z_1^{D+1},\cdots,\bm z_K^{D+1}]\in \mathbb{R}^{2M\times K},\\
    &\bm Z_w = \sqrt{P_d} \frac{\bm Z_w}{\|\bm Z_w \|_F },\\
    &\bm W= \bm Z_w(1:M,:) +j\bm Z_w(M+1:2M,:),
\end{align}
where the notation $\bm Z(i_1:i_2,:) $ denotes the submatrix of $\bm Z$ constructed by taking the rows of $\bm Z$ indexed from $i_1$ to $i_2$.

Note that throughout the GNN architecture, we use the same
$f_w^{0}(\cdot)$, $f_0^{d}(\cdot)$, $f_1^{d}(\cdot)$, $f_2^{d}(\cdot)$,
$f_3^{d}(\cdot)$, $f_4^{d}(\cdot)$, and $f_w^{D+1}(\cdot)$
to update the node representation vectors for all the user nodes. This allows
the GNN to be generalizable to an IRS network with arbitrary number of users.
The learned combining and aggregation operations \eqref{eq:gnn_update_2} and \eqref{eq:gnn_update_1} are independent of the number of users. If we increase or decrease the number of users in the system, we only need to increase or decrease the number of nodes in the graph, the same learned combining and aggregation operations can still be used without having to re-train the neural network. 
\begin{figure*}[!t]
    \centering
    \includegraphics[width=4.2in]{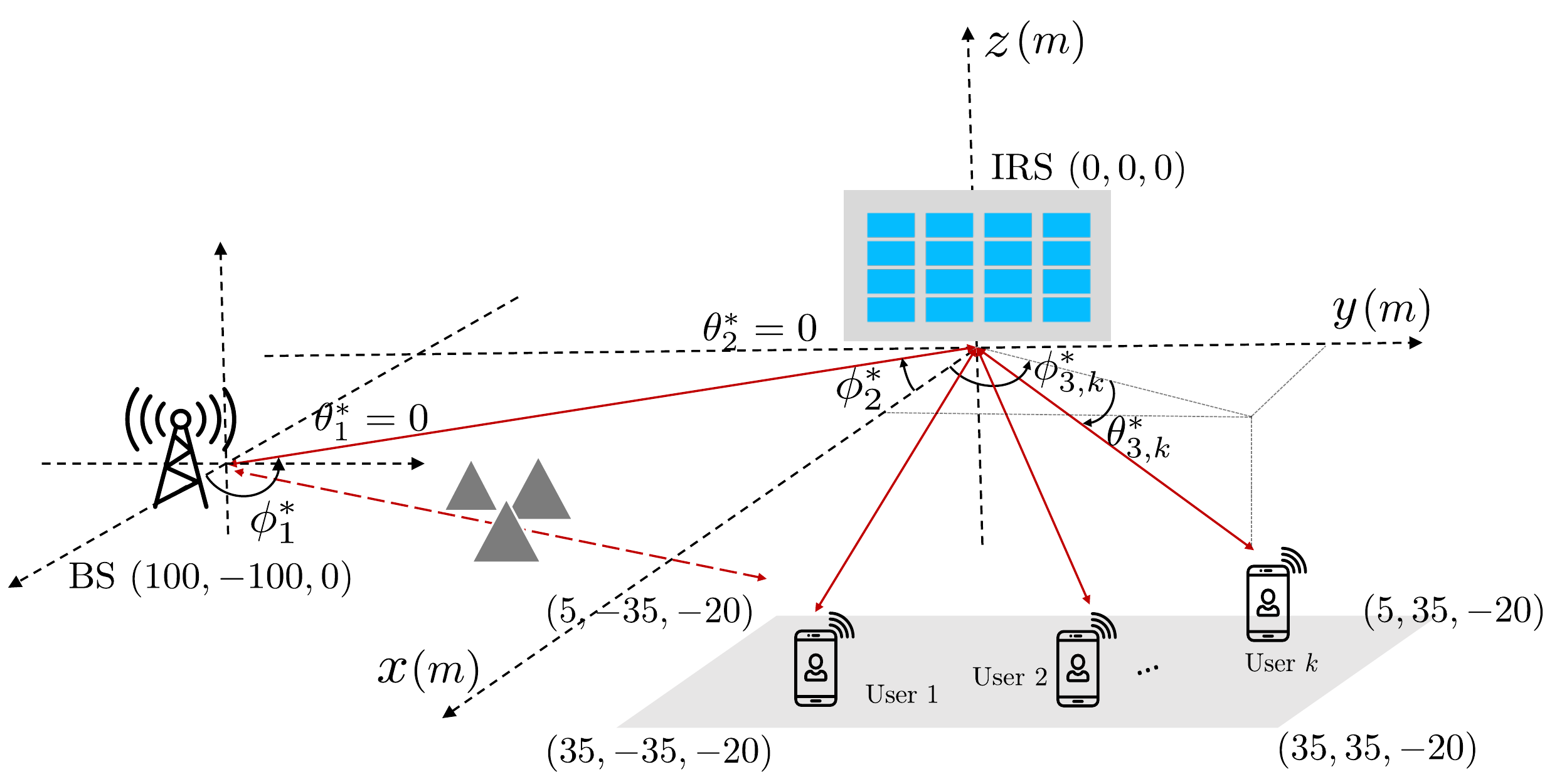}
    \caption{Simulation layout of IRS assisted communication system.}
    \label{fig:simulation_layout}
\end{figure*}
\subsection{Neural Network Training}
Since the existing deep learning software packages do not support complex-valued operations, to compute the network utility during the training phase, we rewrite the achievable rate $R_k$ as a function of the real and imaginary parts of $\bm w_k$ and  $\bm v$ as follows:
\begin{equation}
    R_k = \log\left(1+\frac{\|\bm\gamma_k\|^2}{\sum_{i=1,i\neq k}^K \|\bm\gamma_i\|^2+\sigma_0^2}\right),
\end{equation}
where
\begin{align}
    \setlength\arraycolsep{1pt}
    \bm\gamma_i =&
    \begin{bmatrix}
        \Re\{\bm w^\top_i\}&-\Im\{\bm w^\top_i\}\\
        \Im\{\bm w^\top_i\}&\Re\{\bm w^\top_i\}
    \end{bmatrix}\notag\\
    &\cdot
    \left(
        \begin{bmatrix}
            \Re\{\bm h_k^{\rm d}\}\\
            \Im\{\bm h_k^{\rm d}\}
        \end{bmatrix}+\begin{bmatrix}
            \Re\{\bm A_k\}&-\Im\{\bm A_k\}\\
            \Im\{\bm A_k\}&\Re\{\bm A_k\}
        \end{bmatrix}
        \begin{bmatrix}
            \Re\{\bm v\}\\
            \Im\{\bm v\}
        \end{bmatrix}
    \right).
 \end{align}
Given this real representation of $R_k$, the loss function of the GNN can be expressed as $-\mathbb{E}\left[U( R_1(\bm v,\bm W),\ldots, R_K(\bm v,\bm W) )\right]$. Note that we need CSI to generate training samples and to compute the network utility, but once trained, the operation of the neural network does not require CSI. \re{We remark that the training of the neural network is done offline, so it does not affect the run-time complexity of the proposed approach.}

During training, the neural network learns to adjust its weights to maximize the network utility, i.e., the objective function of problem \eqref{prob:formulation}, in an unsupervised manner, using the stochastic gradient descent method. The updates of the neural network parameters including the computation of  the corresponding gradients can be automatically implemented using any standard numerical deep learning software package.

The overall end-to-end training allows us to jointly design the beamforming matrix and phase shifts from the received pilots directly. As shown in the simulation results in the next two sections, the proposed deep learning method is able to solve problem \eqref{prob:formulation} more efficiently, in the sense that it would need fewer pilots to achieve the same performance as the conventional separated channel estimation and network utility maximization approach.

\section{Performance for Sum-Rate Maximization}
\label{sec:sum_rate}
In this section, we evaluate the performance of the proposed deep learning method for the problem of sum-rate maximization in comparison to the channel estimation based approach. Specifically, the sum-rate maximization problem is given as
\begin{equation}\label{prob:sumrate}
    \begin{aligned}
        & \underset{\begin{subarray}{c}
            (\bm W, \bm v)=g(\bm Y)
        \end{subarray}}{\maxi}&\quad &\mathbb{E}\left[\sum_k R_k(\bm v,\bm W)\right]\\
        & \subj&\quad& \sum_k\mathbb\|\bm w_k\|^2\le P_d,\\
        &&&|v_i| = 1,i=1,2,\cdots,N,
    \end{aligned}
\end{equation}
so the loss function can be set as $-\mathbb{E}\left[\sum_k R_k(\bm v,\bm W)\right] $.

\subsection{Simulation Setting}
We consider an IRS assisted  \re{multiuser MISO} communication system as illustrated in Fig.~\ref{fig:simulation_layout}, consisting of a BS with $8$ antennas and an IRS with $100$ passive elements. As shown in the Fig.~\ref{fig:simulation_layout}, \re{the $(x, y, z)$-coordinates of the BS and the IRS locations in meters are (100,100, 0) and (0, 0, 0), respectively.} There are $3$ users uniformly distributed in a rectangular area $[5,35] \times [-35,35]$ in the $(x,y)$-plane with $z=-20 $ as shown in Fig.~\ref{fig:simulation_layout}. We assume that the IRS is equipped with a uniform rectangular array placed on the $(y,z)$-plane in a $10\times 10$ configuration. 
The BS antennas have a uniform linear array configuration parallel to the $x$-axis.

We assume that the direct channels from the BS to the users follow Rayleigh fading, i.e., 
\begin{align}
    \bm h^{\rm d}_k = \beta_{0,k} \tilde{\bm h}^{\rm d}_k,
\end{align}
where $\tilde{\bm h}^{\rm d}_k \sim\mathcal{CN}(\bm 0,\bm I)$, and $\beta_{0,k}$ denotes the path-loss of the direct link between the BS and the user $k$ modeled (in dB) as $32.6 + 36.7\log(d^{{\rm BU}}_k)$, where $d^{\rm BU}_k$ is the distance of the direct link from the BS to the user $k$.  We assume that the IRS is deployed at a location where a line-of-sight channel exists between the IRS and the users/BS, so we model the channel $\bm h^{\rm r}_k$'s between the IRS and the user $k$ and the channel $\bm G$ between the BS and the IRS as Rician fading channels:
\begin{align}
    &\bm h_k^{\rm r} = \beta_{1,k}\left( \sqrt{\frac{\varepsilon}{1+\varepsilon}} {\tilde{\bm h}}_k^{\rm r, LOS} +\sqrt{\frac{1}{1+\varepsilon} } {\tilde{\bm h}}_k^{\rm r, NLOS} \right),\\
    &\bm G = \beta_2\left( \sqrt{\frac{\varepsilon}{1+\varepsilon}} \tilde{\bm G}^{\rm LOS} +\sqrt{\frac{1}{1+\varepsilon}} \tilde{\bm G}^{\rm NLOS} \right),
\end{align}
where  the superscript ${\rm LOS}$ represents the line-of-sight part of the channel and  the superscript ${\rm NLOS}$ represents non-line-of-sight part, $\varepsilon$ is the Rician factor which is set to be $10$ in simulations, and  $\beta_{1,k}$, $\beta_2$ are  the path-loss from the IRS to the user $k$ and the path-loss from the BS to the IRS, respectively. The entries of $\tilde{\bm G}^{\rm NLOS}$ and $ \tilde{\bm h}_k^{\rm r, NLOS}$ are modeled as i.i.d. standard Gaussian distributions, i.e.,  $[\tilde{\bm G}^{\rm NLOS}]_{ij} \sim\mathcal{CN}( 0,1)$ and  $[\tilde{\bm h}_k^{\rm r, NLOS}]_{i} \sim\mathcal{CN}( 0,1)$. 
The path-losses (in dB) of the BS-IRS link and the IRS-user link are modeled as $30 + 22\log(d^{\rm BI})$ and $30 + 22\log(d^{{\rm IU}}_k)$, respectively, where $d^{\rm BI}$ is the distance between the BS and the IRS, and $d^{{\rm IU}}_k$ is the distance between the IRS and the user $k$ \cite{guo2020weighted,wu2019intelligent}. The uplink pilot transmit power and the downlink data transmit power are set to be $15$dBm and $20$dBm, and the uplink noise power is $-100$dBm and the downlink noise power is $-85$dBm unless otherwise stated.

The line-of-sight part of the channel $\bm h_k^{\rm r}$ is a function of the IRS/user locations. Specifically, let $\phi^{\ast}_{3,k},\theta^{\ast}_{3,k}$ denote the azimuth and elevation angles of arrival
from the user $k$ to the IRS, as shown in Fig.~\ref{fig:simulation_layout}. Then the $n$-th element of the IRS steering vector $\bm a_{\rm IRS}(\phi_{3,k}^{\ast},\theta_{3,k}^{\ast}) $ can be expressed as \cite{9300189}
\begin{multline}
     [\bm a_{\rm IRS}(\phi_{3,k}^{\ast},\theta_{3,k}^{\ast})]_n = \\
     e^{j\frac{2\pi d^{\rm IRS}}{\lambda_c}\{i_1(n)\sin(\phi_{3,k}^{\ast})\cos(\theta_{3,k}^{\ast})+i_2(n) \sin(\theta_{3,k}^{\ast})\}},
\end{multline}
where $d^{\rm IRS}$ is the distance between two adjacent IRS elements, and $\lambda_c$ is the carrier wavelength, $i_1(n)=\mod(n-1,10)$, and $i_2(n) =\lfloor (n-1)/10\rfloor$. In the simulations, we assume $\frac{2d^{\rm IRS}}{\lambda_c}=1$ without loss of generality. Thus, the line-of-sight channel $ \tilde{\bm h}_k^{\rm r, LOS}$ can be written as
\begin{align}
    \tilde{\bm h}_k^{\rm r, LOS}=\bm a_{\rm IRS}(\phi_{3,k}^{\ast},\theta_{3,k}^{\ast}).
\end{align}
Let $(x_k,y_k,z_k)$ denote the location of the user $k$ and $(x^{\rm IRS}, y^{\rm IRS}, z^{\rm IRS})$ denote the location of the IRS,  we have
\begin{align}
    \sin(\phi^{\ast}_{3,k})\cos(\theta^{\ast}_{3,k}) &= \frac{y_k-y^{\rm IRS}}{d^{{\rm IU}}_k},\\
    \sin(\theta^{\ast}_{3,k}) &= \frac{z_k-z^{\rm IRS}}{d^{{\rm IU}}_k},
\end{align}
where $d^{{\rm IU}}_k$ is the distance between the IRS and the user $k$.

Similarly, let $\phi^{\ast}_1,\theta^{\ast}_1$ denote the azimuth and elevation angles of arrival (AOA) to the BS,  then the steering vector of the BS is given by
\begin{align}
    \bm a_{\rm BS}(\phi^{\ast}_1,\theta^{\ast}_1) = [1,\cdots,e^{j\frac{2\pi(M-1)d^{\rm BS}}{\lambda_c} \cos(\phi^{\ast}_1)\cos(\theta^{\ast}_1)}],
\end{align}
where $d^{\rm BS}$ is the distance between two adjacent BS antennas, and we assume $\frac{2d^{\rm BS}}{\lambda_c}=1$. Let $\phi^{\ast}_2,\theta^{\ast}_2$ denote the azimuth and elevation angles of departure (AoD)
from the IRS to the BS, then we can write
\begin{align}
    \tilde{\bm G}^{\rm LOS} = \bm a_{\rm BS}(\phi^{\ast}_1,\theta^{\ast}_1)\bm a_{\rm IRS}(\phi^{\ast}_2,\theta^{\ast}_2)^{\sf H}.
\end{align}
 Given the location of the BS $(x^{\rm BS}, y^{\rm BS}, z^{\rm BS})$ and the location of the IRS $(x^{\rm IRS}, y^{\rm IRS}, z^{\rm IRS})$, we have
\begin{align}
    &\cos(\phi^{\ast}_1)\cos(\theta^{\ast}_1) = \frac{x^{\rm IRS}-x^{\rm BS}}{d^{\rm BI}},\\
    &\sin(\phi^{\ast}_2)\cos(\theta^{\ast}_2) = \frac{y^{\rm BS}-y^{\rm IRS}}{d^{\rm BI}},\\
    &\sin(\theta^{\ast}_2) = \frac{z^{\rm BS}-z^{\rm IRS}}{d^{\rm BI}},
\end{align}
where $d^{\rm BI}$ is the distance between the IRS and the BS.

\begin{figure*}[!t]
    %fig size:(8.5,8.5,4,4)
    \centering
    \subfigure[Sum rate vs. pilot length.]{ \includegraphics[width=3.25in]{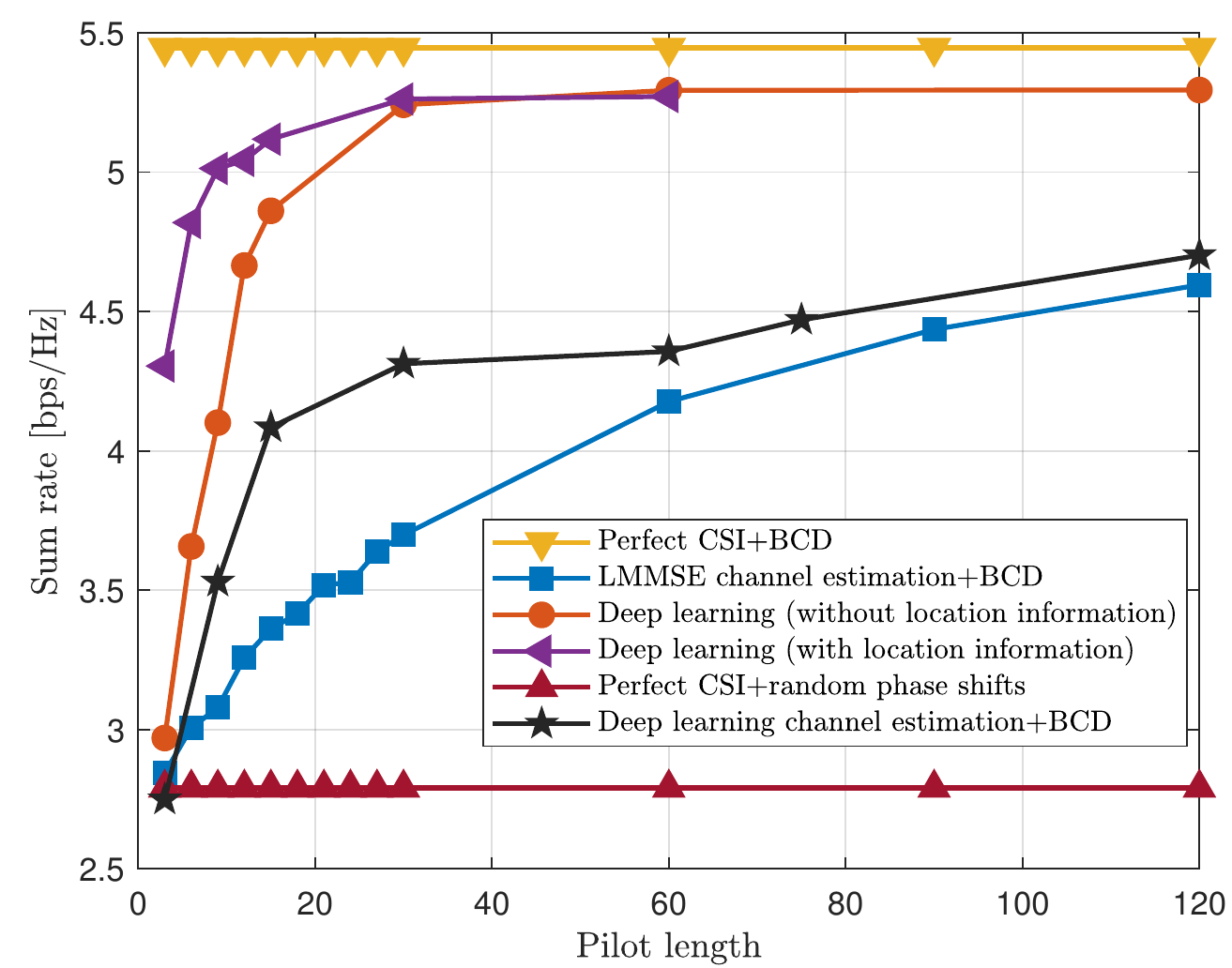}\label{fig:rate_vs_pilot}}
    \subfigure[Testing sum rate vs. training epochs.]{\includegraphics[width=3.1in]{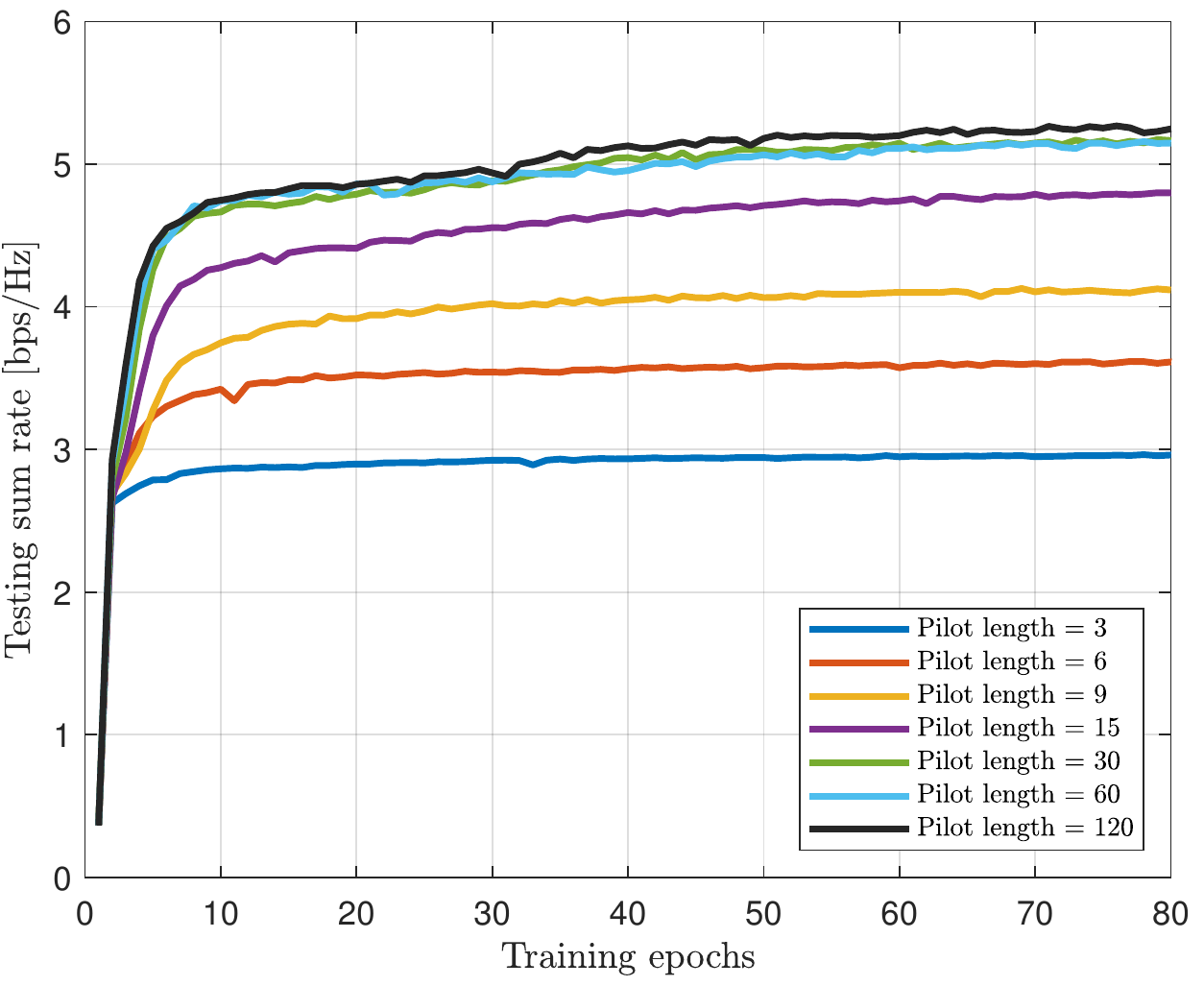}\label{fig:rate_vs_pilot_train}}
    \caption{Performance of the proposed GNN for the IRS system with $M=8 $, $N=100$, $K=3$, $P_d=20$dBm, and $P_u=15$dBm. }
\end{figure*}

\begin{figure*}[t]
    %fig size:(8.5,8.5,4,4)
    \centering
    \subfigure[Sum rate vs. downlink transmit power for $K=3$ and $P_u=15$dBm.]{ \includegraphics[width=2.3in]{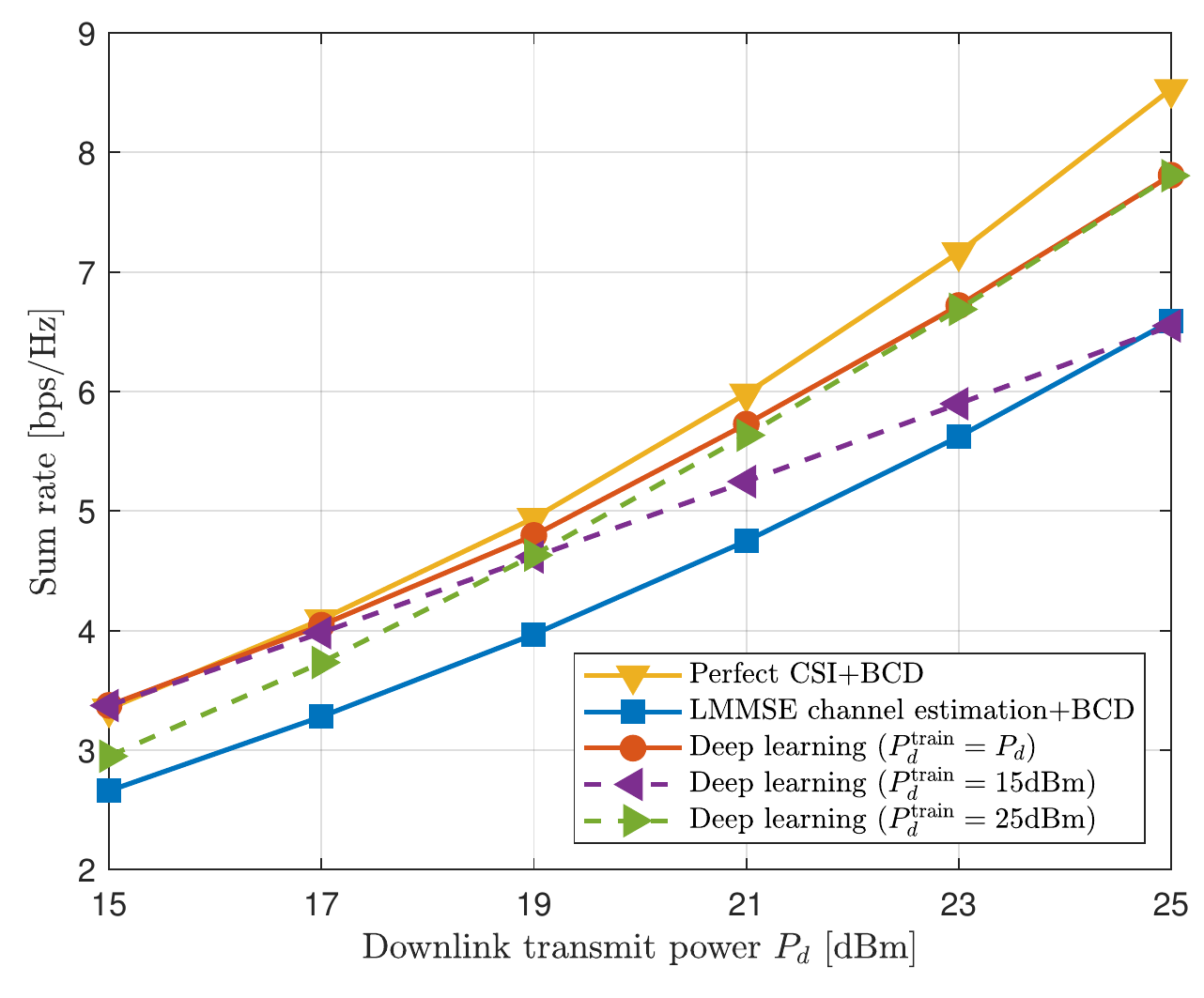}\label{fig:rate_Pd}}
    \subfigure[Sum rate vs. uplink pilot transmit power for $K=3$, $P_d=20$dBm.]{\includegraphics[width=2.3in]{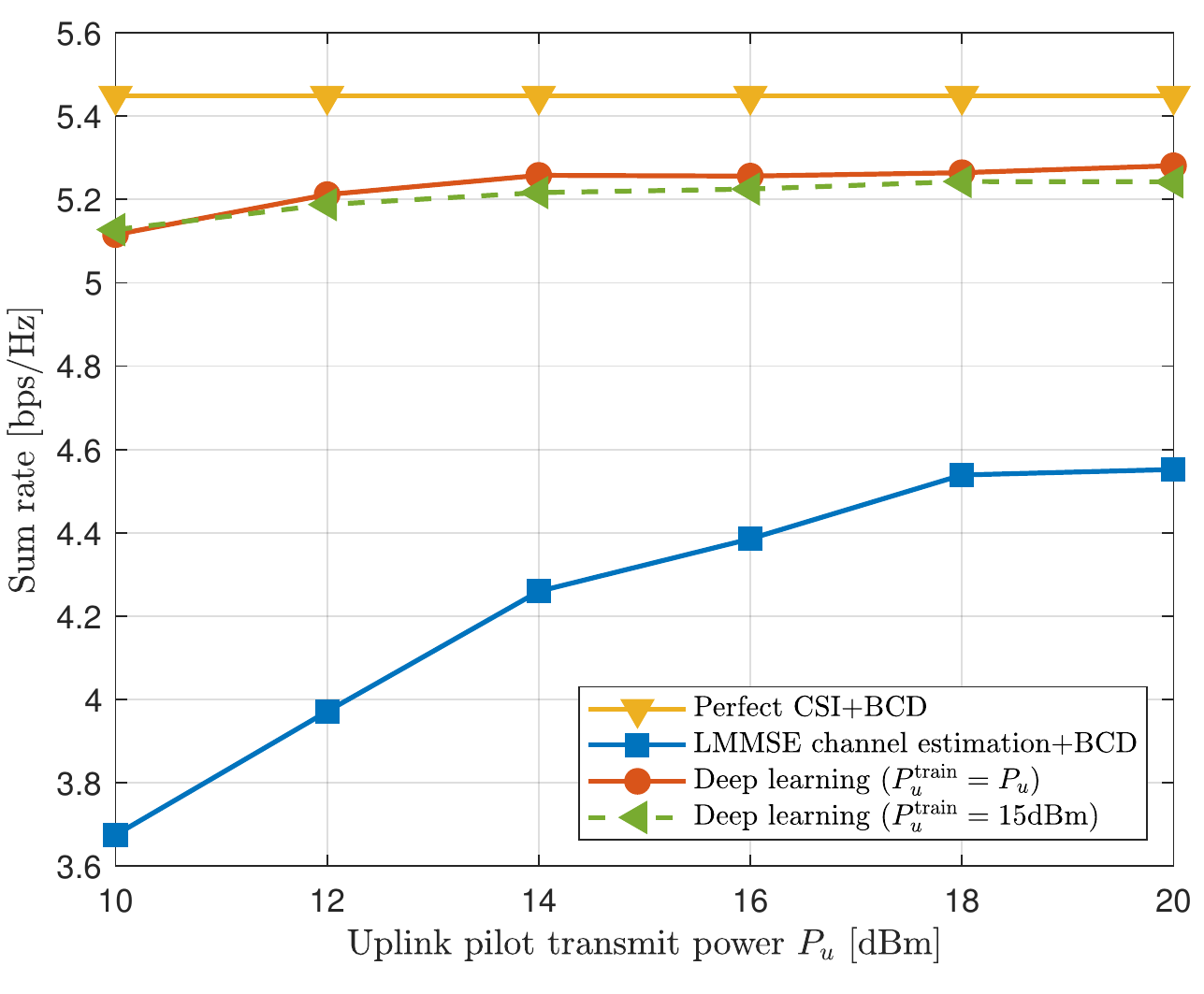}\label{fig:rate_Pu}}
    \subfigure[Sum rate vs. number of users for $P_u=15$dBm and $P_d=25$dBm.]{\includegraphics[width=2.4in]{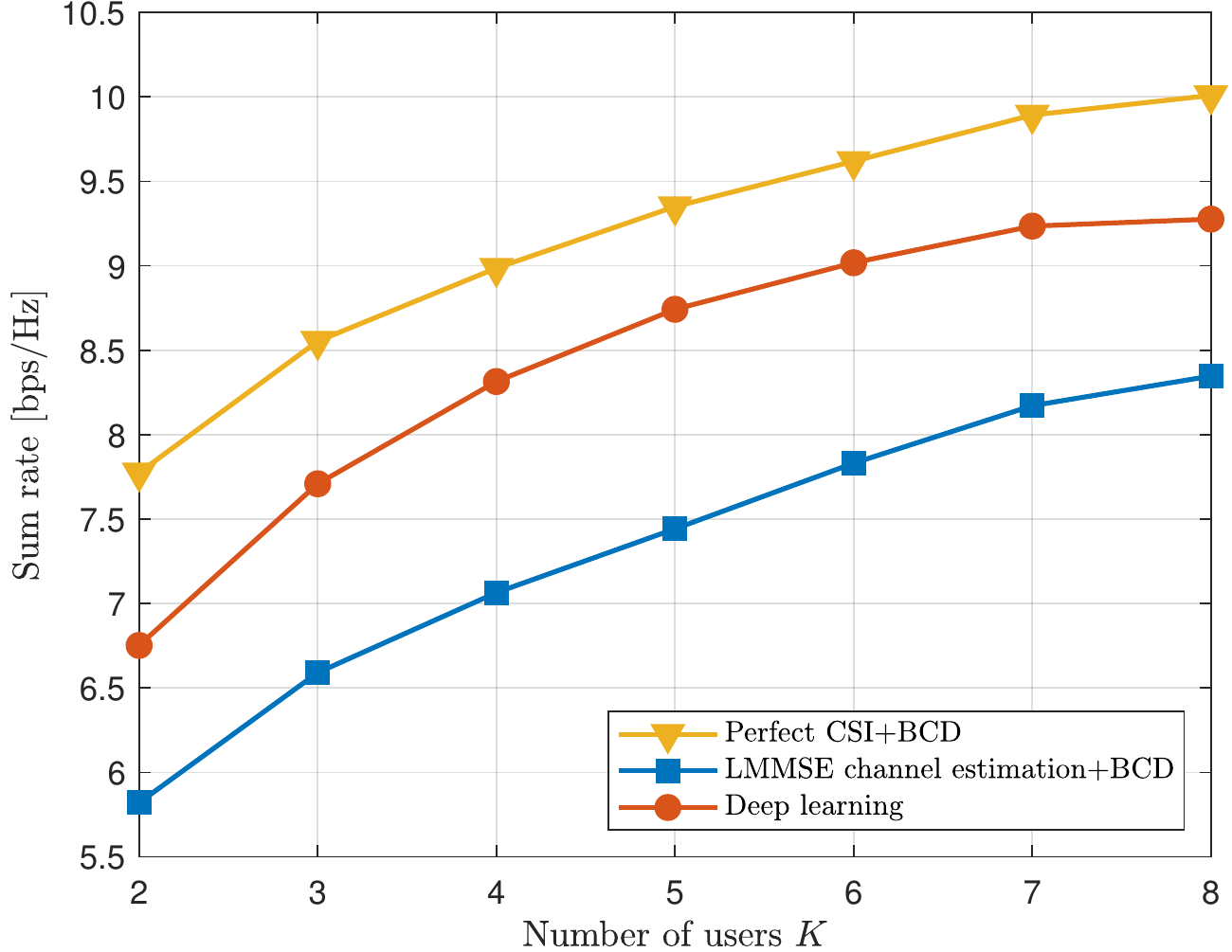}\label{fig:sumrate_K}}
    \caption{Generalization performance of the GNN in an IRS system with $M=8 $, $N=100$ and $L=25K$, where $K$ is the number of users.}
\end{figure*}

\subsection{Neural Network Training and Testing}
We use a 2-layer GNN as proposed in Section \ref{sec:deep_learning_framework}, i.e., with $D=2$. The parameters of the fully connected neural networks $f_w^0$, $f_v^0$ and $f_k^{d}$ with $k=0,\cdots,4, ~d=1,2$  are summarized in Table \ref{table:fcn}.
\begin{table}[t]
    \centering
    \caption{Architecture of the fully connected neural networks.}
    \begin{tabular}{ccc}
    \toprule
    Name  & Size & Activation Function\\
    \midrule
    $f_w^0$&$2M\tau\times 1024\times 512$ &relu\\
    $f_v^0$&$512\times 1024\times 512$ &relu\\
    $ f_0^{1},f_1^{1},f_2^{1},f_3^{1},f_4^{1}$&$512\times 512\times 512$ &relu\\
    $ f_0^{2},f_1^{2},f_2^{2},f_3^{2},f_4^{2}$&$512\times 512\times 512$ &relu\\
    \bottomrule
    \end{tabular}
    \label{table:fcn}
\end{table}

We implement the proposed network using the deep learning library TensorFlow \cite{tensorflow2015-whitepaper}. The neural network is trained  using  the Adam optimizer  \cite{kingma2014adam} with an initial learning rate $10^{-3}$, and the learning rate is decreased after $300$ iterations by a factor of $0.98$. At each training epoch, we iterate $100$ times to update the parameters of the neural network, and $1024$ training samples are used to compute the gradients in each iteration. We terminate the training process if the loss function does not decrease on the validation data set over $10$ consecutive training epochs.

In the testing stage, we compare the neural network with the following benchmarks:
\begin{itemize}
    \item \textit{Benchmark 1. Perfect CSI with BCD}: Given perfect CSI, the sum-rate maximization problem is solved by the block coordinate descent (BCD) algorithm proposed in \cite{guo2020weighted}. We stop the BCD algorithm when the increase in sum rate between two consecutive iterations is below $10^{-3}$.
    \item \textit{Benchmark 2. LMMSE channel estimation with BCD}: We first estimate the channels using the LMMSE estimator developed in Section \ref{sec:baseline}, then perform sum-rate maximization using the BCD algorithm \cite{guo2020weighted}. The required statistics for the LMMSE estimator are computed from $10,000$ channel realizations.
    \item \textit{Benchmark 3. Deep learning based channel estimation with BCD}: To understand the benefit of implicit vs. explicit channel estimation, we implement a neural network to explicitly estimate all the channels followed by using BCD for designing the phase shifts and the beamformers, and compare its performance with the implicit channel estimation strategy proposed in this paper. Specifically, the neural network architecture is almost the same as the proposed GNN for the sum-rate maximization problem, except that the graph node $k=0$ is removed since we only have $K$ channel matrices to be estimated. For the graph node $k\neq 0$, the last layer representation vector $\bm z_k^{D}$ is fed to a linear layer of size $2M(N+1)$ to output the vectorized channel matrix estimation $\bm{\hat{F}}_k$. The inputs to the GNN for the channel estimation are also $\bm{\tilde Y}_1,\cdots, \bm{\tilde Y}_K$, and the loss function for channel estimation is the averaged MSE of the channels over all the users, i.e., $\frac{1}{K}\sum_k \mathbb{E}[\|  \bm{\hat{F}}_k -\bm F_k\|_F^2]$. The implementation of this neural network is the same as the proposed GNN for solving the sum-rate maximization problem \eqref{prob:sumrate} but with a different initial learning rate $10^{-4}$.
\end{itemize}

\subsection{Numerical Results}
We first illustrate the impact of uplink pilot length on the  downlink sum rate in Fig.~\ref{fig:rate_vs_pilot}. All the points are averaged over $1000$ channel realizations. As a trivial baseline, we also show the performance of the system with random phase shifts, while the beamforming matrix $\bm W$ is optimized using the WMMSE algorithm \cite{shi2011iteratively} assuming perfect CSI, to see how much gain we can obtain by jointly optimizing the phase shifts and the beamforming matrix.

From Fig.~\ref{fig:rate_vs_pilot}, using only $30$ pilots, the proposed deep learning approach without location information can achieve about $96\%$ of the sum rate in Benchmark 1 which assumes perfect CSI. The deep learning approach using $15$ pilots still achieves better performance than the Benchmark 2 with $120$ pilots. For Benchmark 2, $303$ pilot symbols are needed for perfect channel reconstruction in the noiseless case. These observations show that the deep learning approach, which directly optimizes the system objective based on the pilots received, can reduce the pilot training overhead significantly.

Further, Fig.~\ref{fig:rate_vs_pilot} shows that incorporating the user location information to the input of the proposed GNN can further improve the sum rate if we do not have sufficiently large pilot length, as compared to the results without location information. But as the pilot length increases, the improvement becomes marginal. This is because the received pilot signals implicitly contain the location information of the users, so it can be learned by the neural network if the pilot length is sufficiently large.

Fig.~\ref{fig:rate_vs_pilot} illustrates that the Benchmark 3 (Deep learning based channel estimation with BCD) can outperform the conventional LMMSE based channel estimation approach. However, the proposed deep learning approach for directly maximizing the sum rate achieves even better performance, which shows that the neural network for direct rate maximization is able to extract more pertinent information from the received pilots than the neural network for explicit estimation of the channel matrix $\hat{\bm F_k}$. Note also that the dimension of the output of the neural network for solving the sum-rate maximization problem is much smaller than the output of the neural network needed for channel estimation. This is another reason that it is advantageous to maximize the sum rate directly instead of recovering the entries of the channel matrix first.

Next, we show how much data is needed to train the proposed neural network. In Fig. \ref{fig:rate_vs_pilot_train}, we plot the sum rate evaluated on testing data against the training epochs for the deep learning approach without location information. Recall that we sample $102,400$ training data in each training epoch. It can be seen from Fig. \ref{fig:rate_vs_pilot_train} that $10$ training epochs are sufficient to achieve a satisfactory performance (greater than $90\% $ of the sum rate achieved with $80$ training epochs).

\re{To evaluate the performance of GNN when the BS is equipped with larger number 
of antennas, we train the same GNN under the setting $N=100$, $K=3$, $P_d=25$dBm, $P_u=15$dBm, for both $M=8$ and $M=16$. As can be seen from Table \ref{table:large_M}, the proposed GNN always outperforms the benchmark of LMMSE with BCD. As expected the gain is larger when the pilot length  $L$ is smaller. But, we also observe that the gain is smaller when the number of antennas $M$ increases from 8 to 16. This is because the problem dimension increases as the number of BS antennas increases. We would need to increase the number of parameters in the GNN to further improve its performance. }

\begin{table}[!t]
    \centering
    \re{\caption{Sum rate ({\rm bps/Hz}) with {\rm $N=100$, $K=3$, $P_d=25$dBm, $P_u=15$dBm}.}
    \begin{tabular}{|c|c|c|c|c|c|}
        \hline $M$ & $L$ & Deep Learning & LMMSE+BCD & Gain & Perfect CSI \\
        \hline \multirow{2}{*} {8} & 45  & $7.45$ & $5.83$ & 28\% & \multirow{2}{*}{8.5} \\
        \cline { 2 - 5 } & 75  & $7.81$ & $6.59$ & 18\% & \\
        \hline \multirow{2}{*} {16} & 45 & $9.02$ & $7.76$ & 16\% & \multirow{2}{*}{11.6} \\
        \cline { 2 - 5 } & 75 & $9.70$ & $8.86$ & 9\%  & \\
        \hline
    \end{tabular}}
    \label{table:large_M}
\end{table}

\subsection{Generalizability}

We now test the trained neural network on different system parameters to show its generalization capability.

\subsubsection{Generalization to different downlink transmit powers}
We fix the number of pilot length $L=75$, and train two neural networks under the settings  with downlink transmit power $15$dBm and $25$dBm, respectively. Then the trained neural networks are tested under different downlink transmit powers.  The results are shown in Fig.~\ref{fig:rate_Pd}. For comparison, we also plot the performance of the neural network with the same training and testing downlink transmit power.  All the points are averaged over $1000$ channel realizations.  From  Fig. \ref{fig:rate_Pd}, it can be observed that the proposed deep learning approach can significantly outperform Benchmark 2 under the different downlink transmit powers. Furthermore, training the neural network at either $15$dBm or $25$dBm (but testing it at different powers) only incurs small losses.  This suggests that the proposed neural network generalizes well to different downlink transmit powers.

\subsubsection{Generalization to different uplink pilot transmit powers}
In this scenario, we fix the pilot length $L=75$ and the downlink transmit power $P_d= 20$dBm. We first train a neural network under the setting with uplink pilot transmit power $P_u=15$dBm. We then change the pilot transmit power to generate the testing data. The results are shown in Fig.~\ref{fig:rate_Pu}. We also plot the performance of the neural network with the same training and testing uplink transmit power for comparison.  All the points are averaged over $1000$ channel realizations. In Fig.~\ref{fig:rate_Pu}, training the neural network at a fixed pilot power of $15$dBm (but testing at different pilot powers) achieves almost identical performance as training and testing at the same powers, which implies that the proposed neural network generalizes well under different uplink pilot transmit powers. Besides, the proposed deep learning approach is quite robust to the pilot transmit power variation, while we observe a significant performance degradation of the Benchmark 2 as the uplink transmit power decreases.

\subsubsection{Generalization to different number of users}
In this simulation, the pilot length is set to be $L=25K$, where $K$ is the number of users.
The downlink transmit power is set as $P_d=25$dBm. We train the proposed neural
network with $K=6$ and test its performance on the settings with different
number of users. The results are shown in Fig.~\ref{fig:sumrate_K}. It can be observed
that the proposed GNN is able to generalize to different number of users, and it always
outperforms the explicit channel estimation Benchmark 2.

\section{Performance for Maximizing Minimum Rate}\label{sec:maxmin}

\begin{figure}[!t]
    %fig size:(8.5,8.5,4,4)
    \centering
    \includegraphics[width=3.5in]{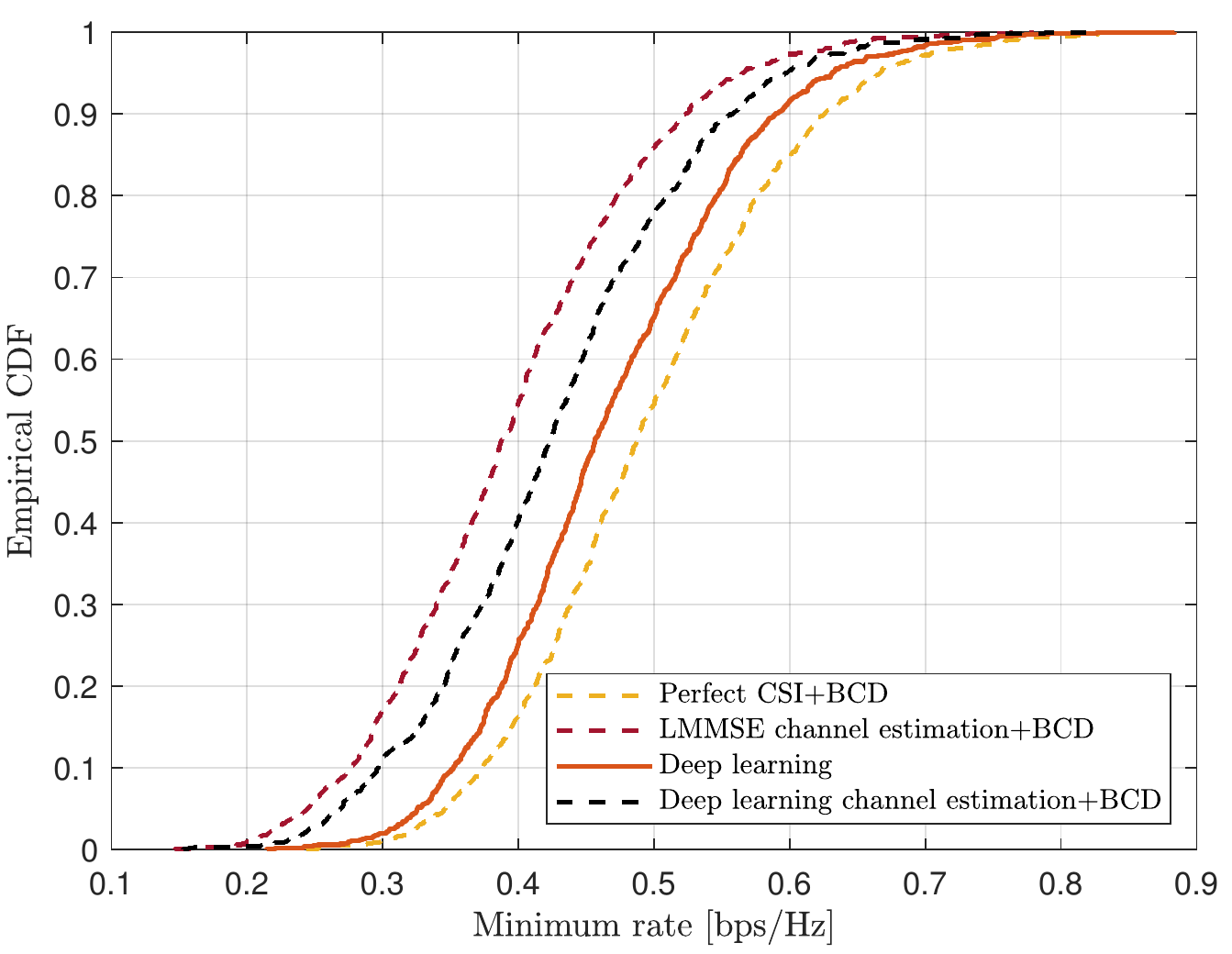}\label{fig:maxmin15}
    \caption{Empirical cumulative distribution function of the minimum user rate with $M=4 $, $N=20$, $K=3$, $L=75$, $P_d=20$dBm, and $P_u=15$dBm.}\label{fig:maxmin}
\end{figure}

The previous section shows the benefits of bypassing explicit channel estimation for the sum-rate maximization problem. But the sum-rate objective does not provide fairness across the users,
which is typically required in practical wireless communication systems. In
this section, we consider the problem of maximizing the minimum user rate,
i.e., max-min problem, as formulated below:
\begin{equation}\label{prob:max-min}
    \begin{aligned}
        & \underset{\begin{subarray}{c}
            (\bm W, \bm v)=g(\bm Y)
        \end{subarray}}{\maxi}&\quad &\mathbb{E}\left[ \min_k R_k(\bm v,\bm W) \right] \\
        & \subj&\quad& \sum_k\mathbb\|\bm w_k\|^2\le P_d, \\
        &&&|v_i| = 1,i=1,2,\cdots,N.
    \end{aligned}
\end{equation}

For the max-min problem \eqref{prob:max-min}, we use the same neural network architecture as in the sum-rate maximization problem, but the loss function is now $-\mathbb{E}\left[ \min_k R_k(\bm v,\bm W) \right] $. We should note that the minimum user rate function is differentiable with respect to $\bm W$ and $\bm v$ almost everywhere except at the points when $R_{k}=R_{j}$ for $k\neq j$, thus the gradient based back-propagation method can still be applied for training the proposed neural network.

The simulation setting is that of an IRS assisted wireless communication system in which a BS with $M=4$ and an IRS with $N=20$ serve $K=3$ users distributed in the rectangular area of $[5,15]\times[-15,15]$ in the $(x,y)$-plane with $z=-20$. The locations of the BS and the IRS remain the same as in Fig.~\ref{fig:simulation_layout}. The downlink transmit power is $20$dBm, and the uplink pilot length is $75$. The training parameters are the same as in the sum-rate maximization problem.

For the baseline with the LMMSE channel estimator, the max-min problem is solved using the BCD algorithm between the beamformer and the reflective coefficient as proposed in \cite{alwazani2020intelligent} in which semidefinite relaxation (SDR) is used for designing the reflective coefficients. The complexity of SDR is high; this is why we choose a simulation setting with $N=20$.

In Fig.~\ref{fig:maxmin}, we plot the empirical cumulative distribution function of the minimum user rate over $1000$ channel realizations. As can be seen from  Fig.~\ref{fig:maxmin}, the proposed deep learning method outperforms the baselines with either the LMMSE or the deep learning channel estimation. This shows that the proposed deep learning framework can also be applied to the max-min problem.

In Table~\ref{table:minrate_K}, we test the generalization capability of the trained GNN for settings with different number of users.  The GNN is trained with $3$ users, but is also tested in the case with $2$ users and $4$ users. It can be observed that the deep learning approach always achieves $87\%-90\%$ of the perfect CSI benchmark for $L=25K$ (where $K$ is the number of users) in all cases, which shows that the GNN generalizes well to settings with different number of users. We also observe that the deep learning approach with $5K$ pilot length can achieve over $94\%$ of the minimum rate achieved by the conventional LMMSE approach with $25K$ pilot length, which shows the remarkable pilot length reduction achieved by the proposed GNN for the max-min problem.

\begin{table}[!t]
    \centering
    \caption{Generalizability to number of users: minimum  rate {\rm(bps/Hz)} with {\rm $M=4 $, $N=20$, $P_d=20$dBm, and $P_u=15$dBm}.}
    \begin{tabular}{|c|c|c|c|c|}
        \hline $L$ & $K$ & Deep learning & LMMSE+BCD & Perfect CSI \\
        \hline \multirow{3}{*} {$L=5K$} & 2 & $0.587$ & $0.529$ & $0.786$ \\
        \cline { 2 - 5 } & 3 & $0.386$ & $0.335$ & $0.496$ \\
        \cline { 2 - 5 } & 4 & $0.274$ & $0.240$ & $0.351$ \\

        \hline \multirow{3}{*} {$L=25K$} & 2 & $0.726$ & $0.620$ & $0.786$ \\
        \cline { 2 - 5 } & 3 & $0.466$ & $0.395$ & $0.496$ \\
        \cline { 2 - 5 } & 4 & $0.315$ & $0.284$ & $0.351$ \\
        \hline
    \end{tabular}
    \label{table:minrate_K}
\end{table}

\section{Interpretation of Solutions from GNN}
\label{sec:interpretation}

\begin{figure*}[!t]
    \centering
    \subfigure[Array response of BS, $M=8$.]{\includegraphics[width=0.24\linewidth]{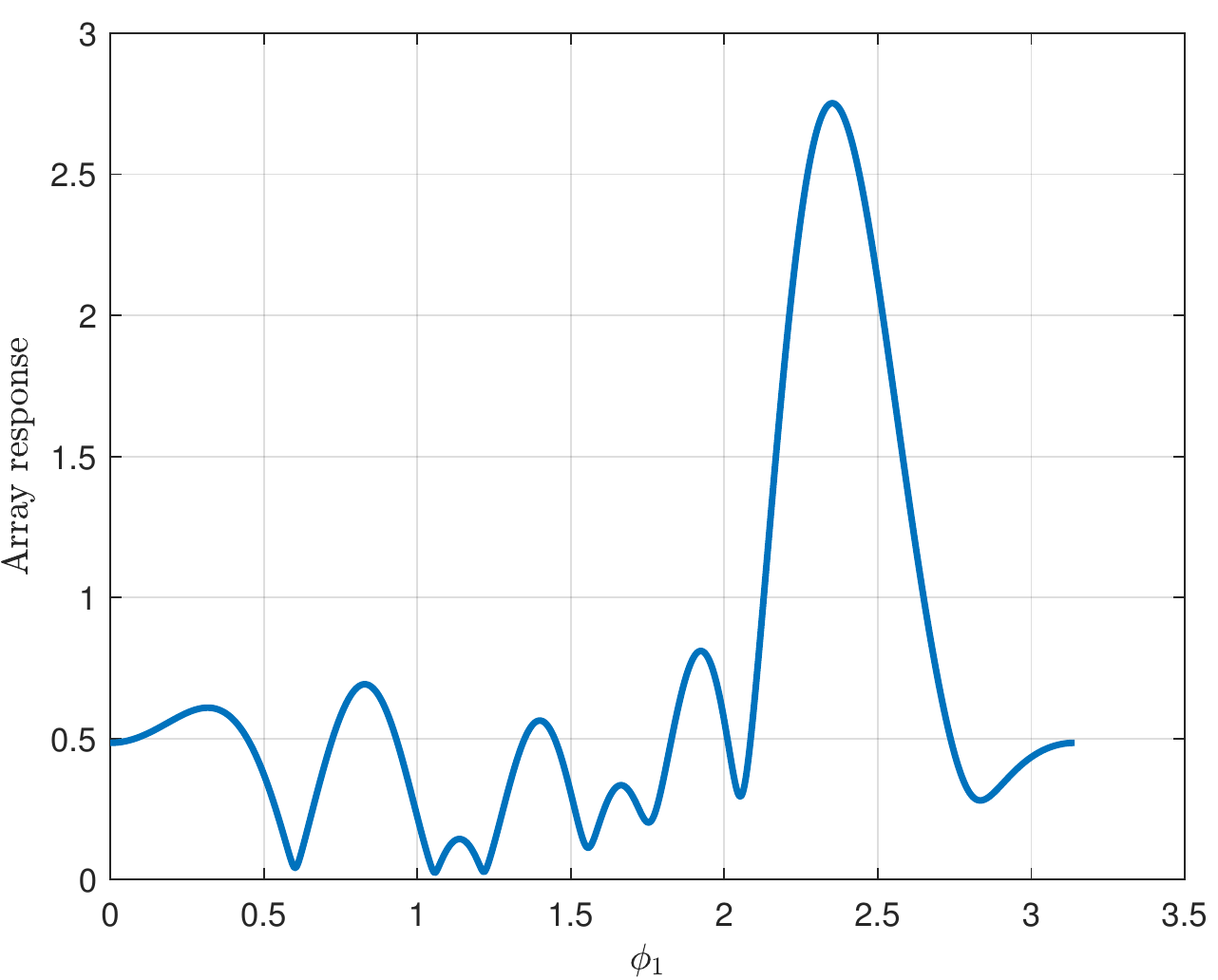}\label{fig:bs_array_respons2}}
    \subfigure[Array response of IRS, $N=100$.]{ \includegraphics[width=0.24\linewidth]{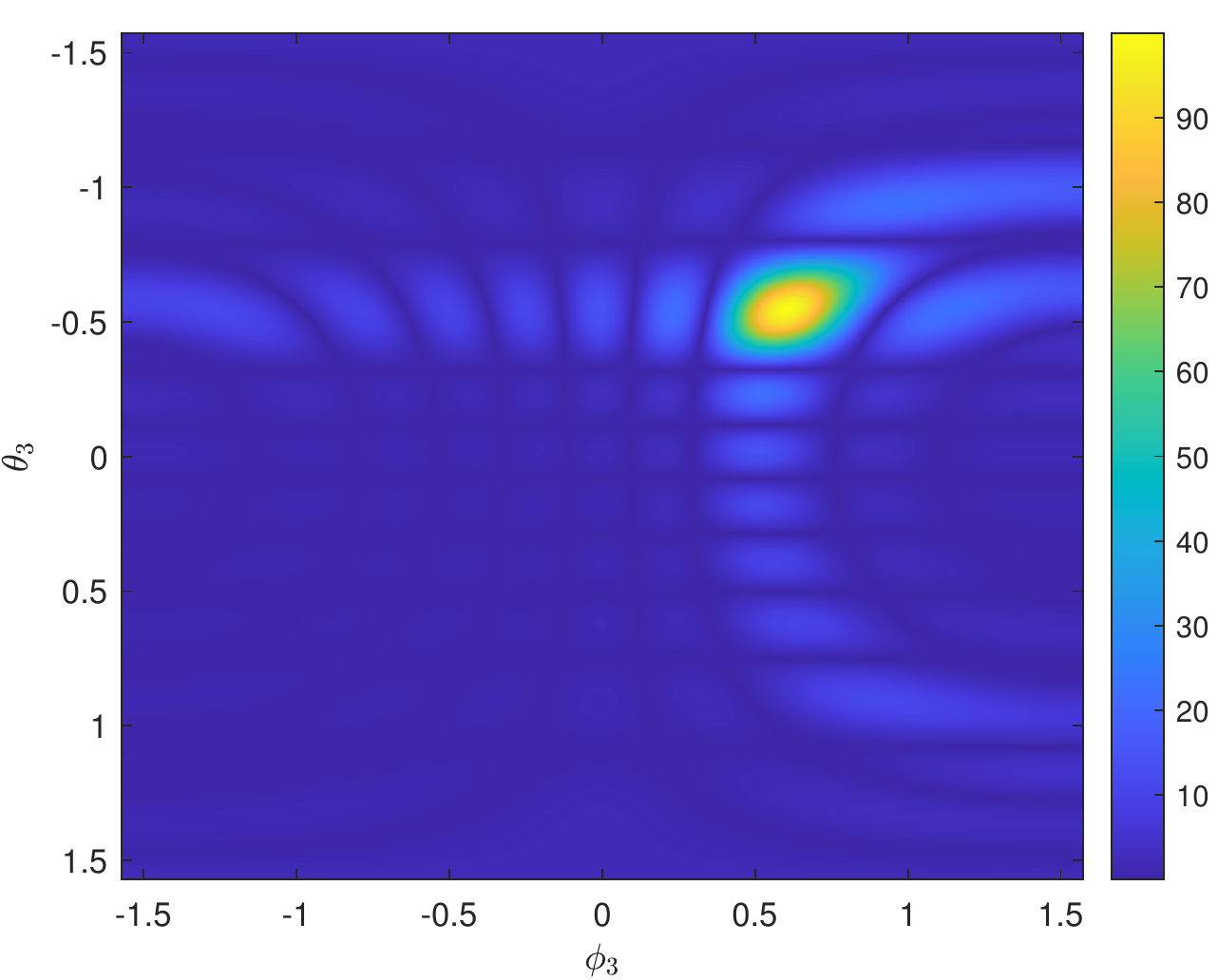}\label{fig:irs_array_respons2}}
    \subfigure[Array response of IRS, $N=50$.]{ \includegraphics[width=0.24\linewidth]{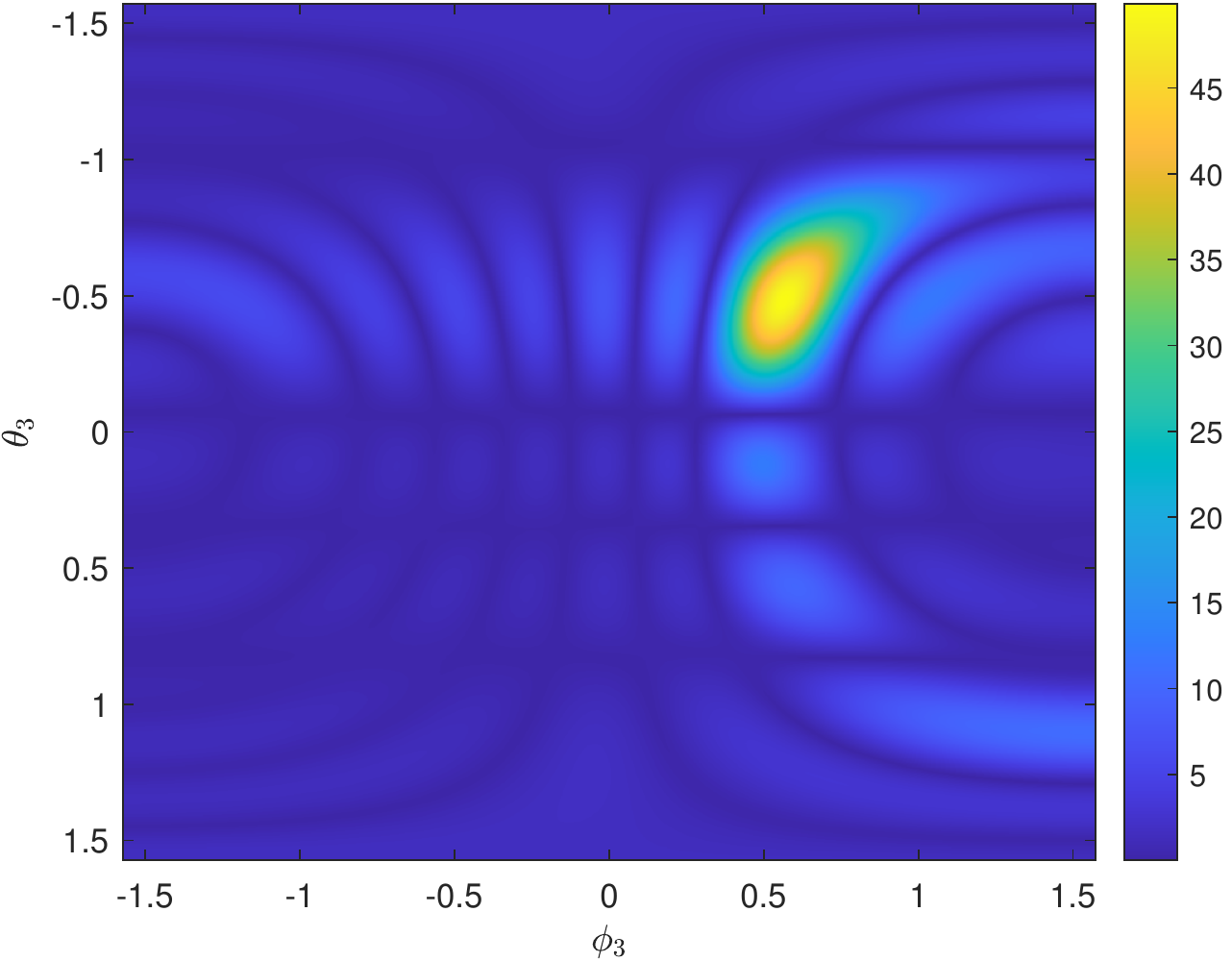}\label{fig:irs_array_respons50}}
    \subfigure[Array response of IRS $N=30$.]{\includegraphics[width=0.24\linewidth]{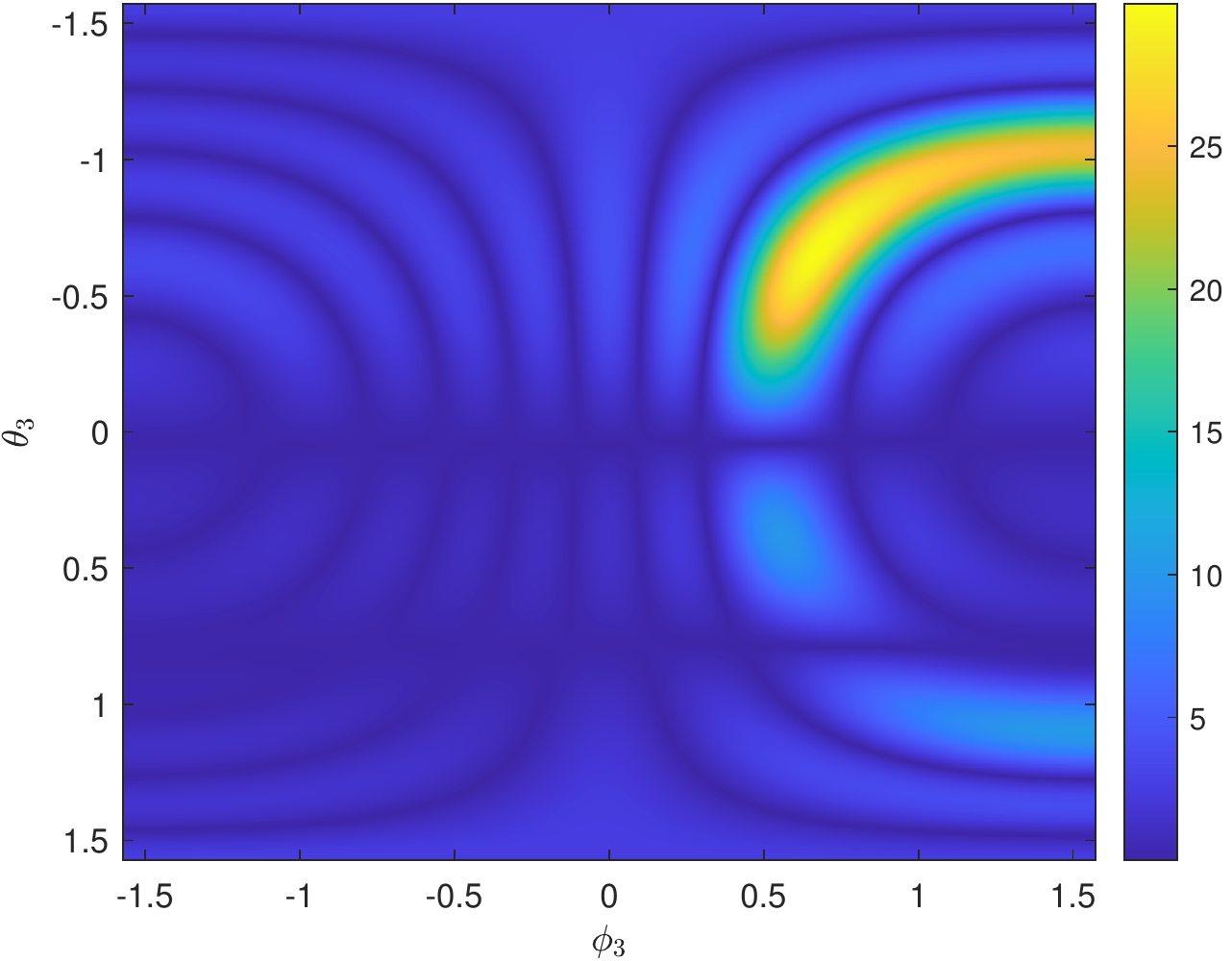}\label{fig:irs_array_respons30}}

    \caption{Array response of the BS and the IRS obtained from GNN for the single-user case: 
(a) and (b) are for $N=100$ and $M=8$; (c) and (d) are for $N=50, 30$ and $M=8$.
The optimal $\phi_1^\ast = 2.356 $, $\phi_3^\ast = 0.588 $, $\theta_3^\ast = -0.506$. 
}
\end{figure*}

\begin{figure*}[!t]
    \centering
    \subfigure[Array response of the BS.]{ \includegraphics[width=0.38\linewidth]{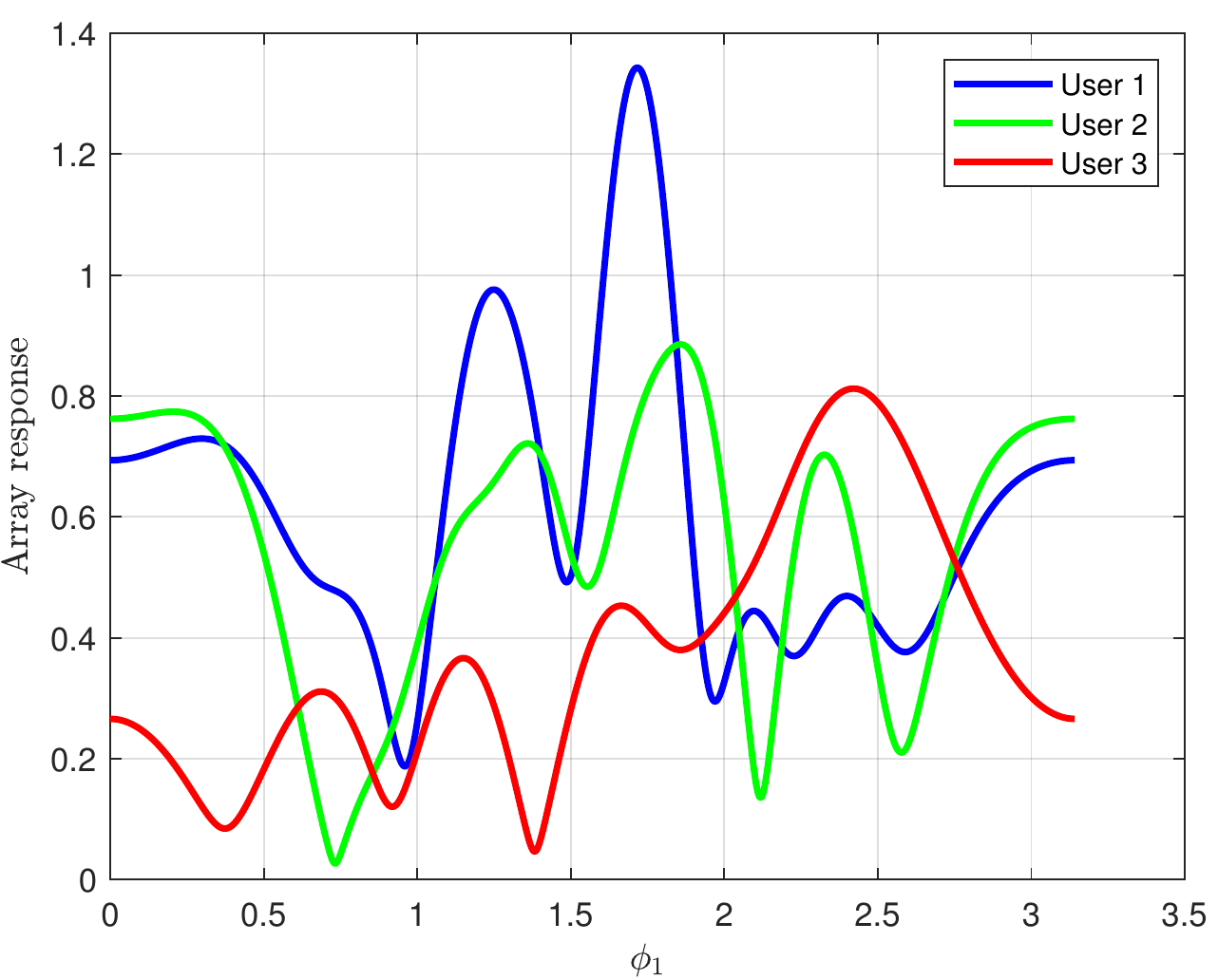}\label{fig:bs_array_respons_mu}}
    \subfigure[Array response of the IRS.]{\includegraphics[width=0.38\linewidth]{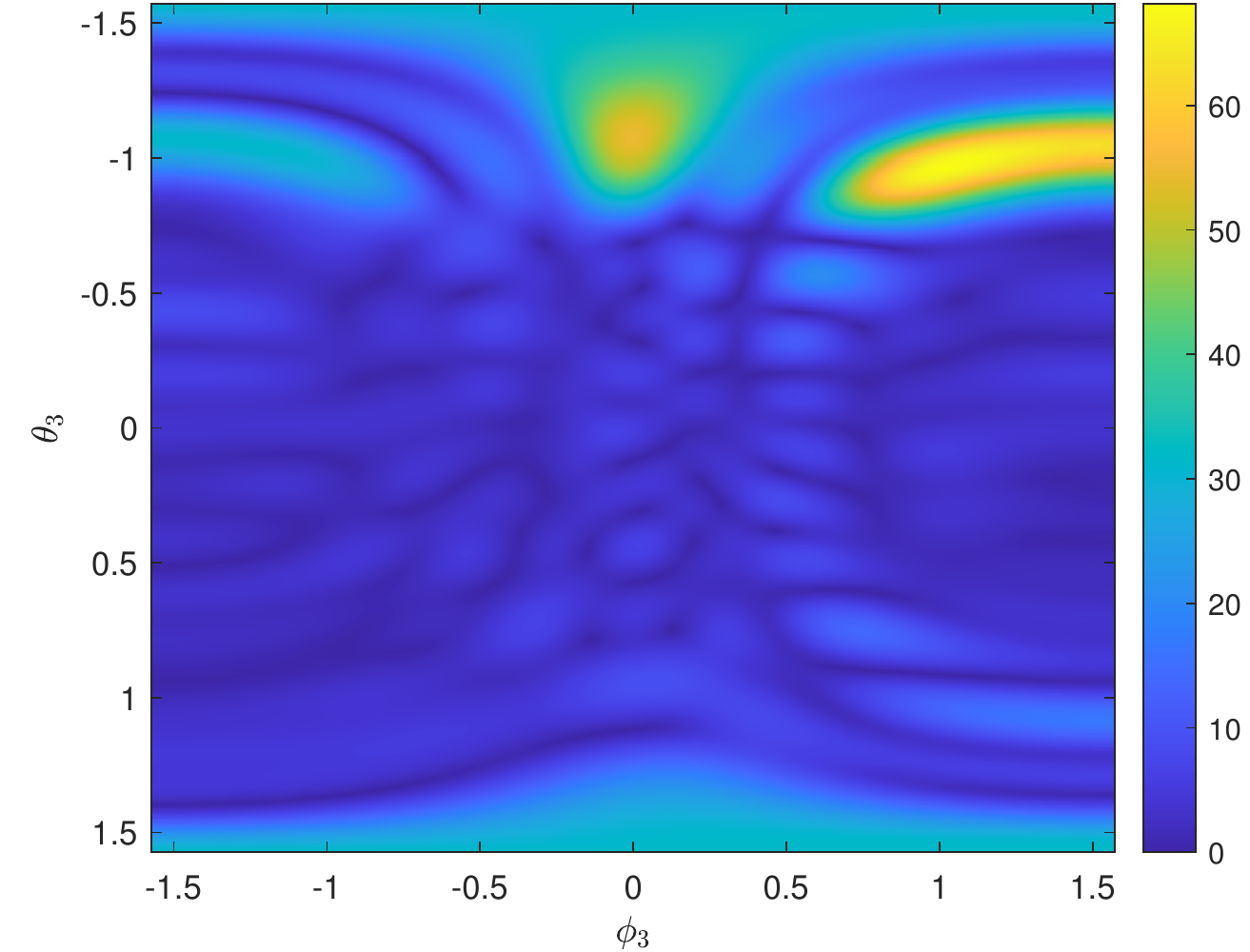}\label{fig:irs_array_respons_mu}}

    \caption{\re{Array response of the BS and the IRS obtained from GNN for maximizing the minimum rate over three users (i.e., $K=3$), with $N=100$, $M=8$. The optimal 
$(\phi_3^\ast,\theta_3^\ast) = (-1.176, -0.994), (0,-0.980), (1.176,-0.994)$ respectively for the three users. The azimuth angle of departure from BS to IRS is $\phi_1^\ast=2.356$.}}
\label{fig:array_response_mu}
\end{figure*}

To interpret the solutions obtained by GNN, in this section, we visually verify
that the learned IRS indeed reflects the signal in the desired directions. We
train the proposed neural network architecture for both a single-user case and \re{a multiuser case}, and use the array response as a way to illustrate qualitatively the correctness of the beamforming pattern. Let $\phi_1, \phi_2, \phi_3$ denote the azimuth angle of arrival from the IRS to the BS,  the azimuth angle of departure from the IRS to
the BS, and the azimuth angle of arrival from the user to the IRS, respectively.
The corresponding elevation angle parameters are denoted by $\theta_1,
\theta_2, \theta_3$, respectively.

The array response at the IRS is a function of the incident angle and the reflection angle of the wireless signals, which is defined as
\begin{align}\label{eq:irs_array_response}
    f_i(\phi_2,\theta_2,\phi_3,\theta_3) &=|\bm a_{\rm IRS}(\phi_2,\theta_2)^{\sf H}\diagg(\bm v)\bm a_{\rm IRS}(\phi_3,\theta_3)|\notag\\
    &=|\bm a_{\rm IRS}(\phi_2,\theta_2)^{\sf H}\diagg(\bm a_{\rm IRS}(\phi_3,\theta_3))\bm v|\notag\\
    % &=|\bm v^{\sf H}\diagg(\bm a_{\rm IRS}(\phi_3))^{\sf H}\bm a_{\rm IRS}(\phi_2)|\\
    &=|\bm v^{\sf H}\bm{\tilde{a}}_{\rm IRS}(\phi_2,\theta_2,\phi_3,\theta_3)|,
\end{align}
where
\begin{align*}
    &[\bm{\tilde{a}}_{\rm IRS}(\phi_2,\theta_2,\phi_3,\theta_3)]_n=\notag\\ &e^{j\pi\{i_1(n)(\sin(\phi_3)\cos(\theta_3)-\sin(\phi_2)\cos(\theta_2))+i_2(n) (\sin(\theta_3)-\sin(\theta_2))\}},
\end{align*}
and $i_1(n)=\mod(n-1,10)$ and $i_2(n) =\lfloor (n-1)/10\rfloor$. 
Similarly, the array response at the BS is a function of the angles $\phi_1,\theta_1$ given by
\begin{align}\label{eq:bs_array_response}
    f_b(\phi_1,\theta_1) &=|\bm a_{\rm BS}(\phi_1,\theta_1)^{\sf H}\bm  w|,
\end{align}
where
\begin{align}
    [\bm{a}_{\rm BS}(\phi_1,\theta_1)]_m=e^{j\pi(m-1)(\cos(\phi_1)\cos(\theta_1))}.
\end{align}
As there exists a line-of-sight component in the channel between the IRS and the BS/user, we expect the learned beamforming vector $\bm w$ and the learned reflection coefficients $\bm v$ to match the angles of the line-of-sight channels, so that the SNR of the user can be maximized.

In the numerical simulation, the pilot length is set to be $L=25K$. 
The locations of the BS and the IRS are $(100, -100,  0)$ and $(0,  0,  0)$ respectively, so that $\phi_1^\ast=2.356, \theta_1^\ast=0$ and $\phi_2^\ast=-0.785, \theta_2^\ast=0$.

We first examine the single-user case, in which, the user is located at $(30, 20, -20)$, 
so that  $\phi_3^\ast= 0.588$ and $\theta_3^\ast=-0.506$.  In Fig.~\ref{fig:bs_array_respons2}, we plot the learned array responses of the BS as a function of $\phi_1$ for the case $N=100$ and $M=8$. It shows that indeed the learned beamforming vector at the BS focuses energy in the direction of the IRS. In Fig.~\ref{fig:irs_array_respons2}, we plot the learned array responses of the IRS as a function of $\phi_3$ and $\theta_3$ (with fixed $\phi_2=\phi_2^\ast$ and $\theta_2=\theta_2^\ast$ according to the BS-to-IRS direction). It is observed that indeed the IRS array response is maximized when $\phi_3 \approx \phi_3^\ast$ and $\theta_3 \approx  \theta_3^\ast$. This shows that the learned configurations of the IRS indeed reflect the signal in the correct user direction.

Moreover, we investigate the impact of the number of elements $N$ at the IRS on
the reflective pattern. For the user at location \re{(30, 20, -20)}, the array
responses of the IRS with $N=30$ and $N=50$ elements (placed as a rectangular
array with 10 elements horizontally)  are shown in
Fig.~\ref{fig:irs_array_respons30} and Fig.~\ref{fig:irs_array_respons50}.
Combined with Fig.~\ref{fig:irs_array_respons2} with
$N=100$, it is clear that the array response focuses better as the number of
elements at the IRS increases.

\re{Next, we examine a multiuser case in which the neural network is trained to maximize the minimum rate over $3$ users located at $(5, -12, -20)$, $(5, 0, -20)$ and $(5, 12, -20)$, so that $(\phi_3^\ast, \theta_3^\ast) = (-1.176, -0.994), (0, -0.980), (1.176, -0.994)$ for the three users respectively. Fig.~\ref{fig:array_response_mu} shows the BS 
and IRS array responses learned by the GNN.  From Fig.~\ref{fig:irs_array_respons_mu},
we indeed see three peaks matching the angles $\phi_3^\ast$ and $\theta_3^\ast$ corresponding to the three users. 
In addition, we observe that the peak corresponding to the first user is weaker than the other two users, but we can see from Fig.~\ref{fig:bs_array_respons_mu} that this user has the strongest array response at the BS. Therefore, the GNN indeed learns to jointly optimize the phase-shifts and the beamforming matrix to ensure fairness in this scenario of maximizing the minimum rate across the three users. 
The weaker IRS array response is compensated by a stronger BS array response.

We also note from Fig.~\ref{fig:bs_array_respons_mu} that the responses of the BS are maximized at different angles for different users.  This shows that the BS beamformers have learned to differentiate the three users. Note that the combined BS and IRS array responses  would need to minimize the interference among the users and not just to maximize each users' direct channels. Note that the Rayleigh channel components from the BS 
to the users also impact the optimum BS and IRS array response patterns. }

\section{Conclusion}
\label{sec:conclusion}
Conventional communication system design always involves obtaining accurate CSI first, then designing the optimal transmission scheme according to the CSI. This design strategy is not practical for IRS due to the large number of passive reflective elements involved. This paper proposes an approach that learns to configure the IRS and beamforming at the BS to maximize the system utility function directly based on the received pilots, in effect bypassing explicit channel estimation. This is accomplished by a generalizable graph neural network architecture that unveils the direct mapping from the received pilots to the desired IRS configuration and the desired per-user beamformers at the BS.  Simulation results show that the trained neural network produces interpretable results and can efficiently learn to solve utility maximization problems using much fewer pilots as compared to the conventional approach. 
 
\bibliographystyle{IEEEtran}
\bibliography{refs}

\end{document}